\documentclass[12pt]{article}

\usepackage{amssymb}

\usepackage{graphicx}
\usepackage{amsmath}
\usepackage{authblk,color,fullpage,gensymb,natbib,rotating,setspace,subfigure,subfig,tabularx,longtable}
\usepackage[english]{babel}

\newcommand{\abs}[1]{\left| #1 \right|}
\newcommand{\E}[1]{\mathbb{E}\left[#1\right]}
\newcommand{\BigO}[1]{\mathcal{O}\left(#1\right)}

\newcommand{\ie}{\textit{i.e.,}\,}
\newcommand{\ba}{\mathbf{a}}
\newcommand{\bsigma}{\boldsymbol \sigma}
\newcommand{\brho}{\boldsymbol \rho}
\newcommand{\bB}{\mathbf{B}}
\newcommand{\bD}{\mathbf{D}}
\newcommand{\bF}{\mathbf{F}}
\newcommand{\bx}{\mathbf{x}}
\newcommand{\bZ}{\mathbf{Z}}

\newcommand{\bc}{\mathbf{c}}
\newcommand{\be}{\mathbf{e}}
\newcommand{\bdf}{\mathbf{f}}
\newcommand{\bv}{\mathbf{v}}

\newcommand{\ip}[2]{( #1 \cdot #2 )}
\newcommand{\R}{\mathbb{R}}
\newcommand{\C}{\mathbb{C}}
\newcommand{\bX}{\mathbf{X}}

\newcommand{\bSigma}{\boldsymbol \Sigma}

\newcommand{\defn}{\buildrel \rm def\over=}

\newcommand{\mat}[1]{\begin{pmatrix} #1 \end{pmatrix}}

\begin{document}



\title{Integrating Theory and Experiment to Explain the Breakdown of Population Synchrony in a Complex Microbial Community}



\author[1,$\dagger$]{Emma J. Bowen}
\author[2,$\dagger$]{Todd L. Parsons}
\author[1]{Thomas P. Curtis}
\author[3]{Joshua B. Plotkin}
\author[4,*]{Christopher Quince}

\affil[1]{Department of Civil Engineering and Geosciences, University of Newcastle, Newcastle-upon-Tyne NE1 7RU, United Kingdom}
\affil[2]{Laboratoire de Probabilit\'{e}s et Mod\`{e}les Al\'{e}atoires, CNRS UMR 7599, Universit\'{e} Pierre et Marie Curie, Paris 75005, France.}
\affil[3]{Department of Biology, University of Pennsylvania, Philadelphia 19104, United States of America}
\affil[4]{Warwick Medical School, University of Warwick, Coventry CV4 7AL, United Kingdom} 

\affil[$\dagger$]{Joint first authors}

\affil[*]{Corresponding author: School of Engineering, Rankine building, Oakfield Avenue, University of Glasgow, Glasgow G12 8LT, United Kingdom; Tel - +441413306458; Fax - +441413304885}
\date{}
\maketitle

\noindent E-mails: EJB - emma.bowen@newcastle.ac.uk; TLP - todd.parsons@upmc.fr;\\
 TPC - tom.curtis@newcastle.ac.uk; JBP - jplotkin@sas.upenn.edu; CQ - c.quince@warwick.ac.uk
\bigskip

\clearpage

\begin{abstract}
We consider the extension of the `Moran effect', where correlated noise generates synchrony between isolated single species populations, to the study of synchrony between populations embedded in multi-species communities. In laboratory experiments on complex microbial communities, comprising both predators (protozoa) and prey (bacteria), we observe synchrony in abundances between isolated replicates.  A breakdown in synchrony occurs for both predator and prey as the reactor dilution rate increases, which corresponds to both an increased rate of input of external resources and an increased effective mortality though washout. The breakdown is more rapid, however, for the lower trophic level. We can explain this phenomenon using a mathematical framework for determining synchrony between populations in multi-species communities at equilibrium. We assume that there are multiple sources of environmental noise with different degrees of correlation that affect the individual species population dynamics differently. The deterministic dynamics can then influence the degree of synchrony between species in different communities. In the case of a stable equilibrium community synchrony is controlled by the eigenvalue with smallest negative real part. Intuitively fluctuations are minimally damped in this direction. We show that the experimental observations are consistent with this framework but only for multiplicative noise.
\bigskip

\end{abstract}

\section{Introduction}
\label{Introduction}
In nature, distinct geographically separated populations of similar species composition often exhibit correlated population dynamics \citep{Grenfell1998}. Expounding the origin and breakdown of such population synchrony is critical to predicting spatial population dynamics \citep{Cazelles2001,Koenig1999}, and understanding metapopulation persistence \citep{Palmqvist1998,Earn2000}. Dispersal and migration play an important role in synchronising connected populations over short distances \citep{Jansen1999}, but when these mechanisms are limited, for example, when populations are distributed over larger geographical areas or  physically isolated,  the `Moran Effect' is often invoked to explain synchronised dynamics \citep{Moran1953,Ranta1997}. Moran showed that correlated environmental noise will synchronise populations with similar intrinsic density dependence, the degree of correlation between the populations being equal to the correlation in the noise itself. However, the original result was derived under a very specific set of mathematical conditions: single species experiencing linear dynamics close to equilibrium with additive white noise. 

There have been experimental and observational confirmations of the Moran effect \citep{Grenfell1998,rosenstock11,koenig13} but these tend to focus on one or a few species, synchrony in complex multi-species has received much less attention. Reactors containing microbial communities provide a convenient tool to study population synchrony in real ecosystems under well-defined conditions and within experimentally realisable timescales \citep{Benton2007}. Further, laboratory reactor communities are useful specifically for examining the Moran effect because they can be composed from the same source community, hence have the same initial population structure; be subjected to the same environmental conditions; and finally, they can be physically isolated, eliminating dispersal and migration.  Thus, they are similar to `island populations' as studied by \citet{Grenfell1998}, for example. 

To explore synchrony in complex communities we ran replicate reactor pairs at four different dilution rates. Each reactor in a pair received the same input substrate but experienced different but correlated environmental noise. Our starting experimental microbial community was sampled from the activated sludge tank of a wastewater treatment plant; representative of a complete, multi-species, functional ecosystem, experiencing similar environmental conditions {\it in situ}, to those we would apply in the laboratory.  From this single source community, we inoculated identical reactors with the same sample volume, these therefore had approximately the same initial population densities of each of the community constituents. Locating the sub-populations in isolated but replicated laboratory reactors provided us with `island' populations as our experimental system.  To examine the occurrence and breakdown of synchrony in our microbial communities, we manipulated the dilution rate of the laboratory reactors. Dilution rate was varied as it is a bifurcation parameter controlling the dynamical regime of the system. We observed a breakdown in synchrony as the dilution rate increased. This is counterintuitive since the reactors are coupled through the substrate and as the rate of input increases naively we might expect the reactor dynamics to converge. To explain this fascinating phenomenon we turned to mathematical theory, considering possible extensions of Moran's theorem. 

There has been a great deal of interest in determining the conditions under which the Moran effect applies \citep{Grenfell1998,Ripa2000,Engen2005}. These studies have mostly focussed on relaxing the non-linear dynamics \citep{Engen2005} or considering populations that are cycling and not at equilibrium \citep{Vasseur2009,Fox2011}. One question that has received much less attention is synchrony in multi-species communities; the degree to which correlations can be transmitted at equilibrium through trophic interactions such as predator-prey or competition.  A theoretical framework has been developed in discrete time for this problem \citep{Greenman2005a,Greenman2005b}.  In the completely general case, in addition to multiple species in a community, there will be multiple independent noise components.  Each of the noise components will impact the different species differently and each can be correlated to different degrees between the communities. Greenman and Benton derived a general framework for this case by linearising the population dynamics about equilibrium and considering the limit of weak noise \citep{Greenman2005a,Greenman2005b}. They applied this to a number of simple models \citep{Greenman2005b}. In general, the degree of correlation between populations is a complex function of the demographic parameters and the noise correlations. This reflects the effect of trophic interactions within the community transmitting the independent noise correlations between trophic groups. 

We adapted Greenman and Benton's approach to a continuous time model. Continuous time models are more appropriate for over-lapping generations and can be more straightforward to analyse. We assume that all species populations are at equilibrium, which is supported by previous experimental data, and that the amplitude of the environmental noise is small so that the dynamics can be linearised. These assumptions are equivalent to the original assumptions of Moran,  as discussed above, the noise has 
multiple independent components and after linearisation each component has a coefficient that is species dependent. This reflects the different ways that the noise can impact each species population dynamics and their differing sensitivity to the components. We can imagine these components representing different aspects of the environment, for example temperature or pH. 

We considered the case of a simple predator-prey chemostat model in detail. To provide generality, we considered alternative sets of modelling assumptions; four are considered in the appendices.   The one which incorporates endogenous predator
 mortality and multiplicative noise had the closest correspondence to the experimental results.  
 Intuitively, synchrony is controlled by the eigenvalue with smallest negative real part and the 
 corresponding eigenvector of the matrix of coefficients of the linearised per capita growth rates. 
 The fluctuations will be minimally damped in this direction. Projecting the noise components on to 
 this eigenvector selects different components, and the correlation of those selected components 
 controls the synchrony. As dilution rate changes the eigenvector that controls the systems switches 
 from being dominated by the predator to being dominated by the prey. If we assume that there are two 
 noise components that differ in their degree of correlation, and that the the prey dynamics are more 
 strongly influenced by the less correlated noise, then this explains the breakdown of synchrony in the reactors.  

\section{Reactor Experiments and Results}

We ran replicate pairs of isolated sequencing batch reactors next to each other with the same input 
substrates and under the same operating conditions. These replicates consequently experienced identical 
resource inputs but different, albeit correlated, environmental noise. Each batch regime took six hours, 
during which time the reactors were filled with substrate, aerated, and mixed for five hours and fifteen 
minutes. Towards the end of this period, a proportion of the mixed reactor liquor, including micro-organisms, 
was removed, and subsequently the liquor was settled and decanted, leaving the microbial biomass in the reactor 
ready for the addition of fresh substrate to repeat the cycle. The removal of a portion of mixed liquor each cycle 
constituted the reactor dilution.

Four pairs of replicates were run for 61 days, with the proportion of the reactor volume removed every six hours, 
$d$, set at four increasing values: (0.025, 0.035, 0.0833, 0.25). This parameter, the dilution fraction, corresponds
 to the dilution rate, $D$, in a continuous flow chemostat with $D$ (hours\textsuperscript{-1}) $= d/6$. These dilution 
 rates were chosen because they correspond to a Solid Retention Time (SRT), \textit{i.e.}, the mean time a particle 
 would expect to spend in the reactors, of 10 days, 7 days, 3 days, and 1 day respectively. The SRT is the typical 
 method of expressing dilution rate in wastewater treatment. In Table \ref{ONE} we give means, standard deviations, 
 and un-lagged (\textit{i.e.}, comparing biomass at the same time points) cross-correlations between replicate 
 pairs for the time series of temperatures experienced by the reactors.  We also give the details of the substrate 
 concentrations in terms of chemical oxygen demand (COD) and total nitrogen concentration (NH$_4$-N). Four reactors 
 were used, and run in two experiments, each of two replicates at two different dilutions, first $d = 0.025$ and $d = 0.25$ 
 together, and then $d = 0.035$ with $d = 0.0833$.  Influent substrate to all four reactors was provided from a single bulk 
 source in each experiment.

Samples of the mixed liquor from the reactors were collected every day for the $d = 0.025$ and $d = 0.25$ reactors, and every other day for the $d = 0.035$ and $d = 0.0833$ reactors.  Microbial biomass was quantified using the standard mixed liquor volatile suspended solids method \citep{APHA1985}: resulting biomass concentrations are shown in Fig. 1. It is apparent from these time series that microbial biomass decreases with increased dilution, as would be expected, but also that the replicates become less synchronised as $d$ increases, a result that is dramatically confirmed by the cross-correlations between replicates plotted as a function of $d$ in Fig. 2. 

Strictly, the mixed liquor biomass is a measurement of all microorganisms, including bacteria and protozoa. However, the bacteria as the lower trophic level will represent the bulk of this biomass. To investigate synchrony amongst protozoa these were preserved, stained and enumerated under light microscopy \citep{Widdicombe2002}. The same breakdown in synchrony was also observed for protozoa counts although the results were not as dramatic. Consequently, any impact of the protozoa component on the observed microbial biomass will be to increase the observed correlation. Therefore, we can conclude that at the three higher dilution rates the protozoa were more correlated than the bacteria (Fig. 2).

The breakdown in synchrony in the experimental reactors with increased dilution cannot simply be explained in terms of the environmental conditions (Table \ref{ONE}). There is no significant relationship between the correlation in temperature and the biomass correlations; and the influent nutrient concentrations were identical across replicates. We therefore hypothesise that the breakdown in synchrony must be associated with changes in the underlying predator-prey dynamics of the system. Predator-prey interactions between bacteria and protozoa have been shown to be very important in reactors of this type \citep{Hughes1976}. This then motivated a full analytical mathematical treatment of synchrony between predator-prey communities in chemostats experiencing correlated noise.

\section{Mathematical Analysis}

We modelled the predator-prey dynamics via the `double Monod chemostat model', 
which describes a system in which continuously input resource (nutrient) is taken up and converted 
into growth by microbial prey organisms, at a rate that is a saturating Monod function of the resource concentration; 
similarly the prey is then consumed by a microbial predator also with Monod uptake. It is unrealistic to assume that in the
 absence of prey and with zero washout that the predator will be maintained indefinitely, 
 so we added an endogenous metabolic term for the predator as described in \citet{Nisbet1983}; 
 this model is widely used as a minimal microbial predator-prey model. In our case, the
  two replica systems, $d = 1,2$, experiencing correlated noises in predator and prey obey
\begin{equation}
\begin{aligned}
	dR^{(d)} &=
	\left[(R_{0}-R^{(d)})D - \frac{\mu_{1}}{\gamma_{1}} \frac{R^{(d)}X^{(d)}}{R^{(d)}+k_{1}}\right]\, dt\\
	dX^{(d)} &= \left[\mu_{1}\frac{R^{(d)}X^{(d)}}{R^{(d)}+k_{1}} - DX^{(d)} -
	\frac{\mu_{2}}{\gamma_{2}} \frac{X^{(d)}Y^{(d)}}{X^{(d)}+k_{2}}\right]\, dt\\
	&\qquad +\varepsilon \varsigma_{11} f_{1}\left(R^{(d)},X^{(d)},Y^{(d)}\right) \, dB^{(d)}_{1}(t)+\varepsilon \varsigma_{12} f_{1}\left(R^{(d)},X^{(d)},Y^{(d)}\right) \, dB^{(d)}_{2}(t)\\
	dY^{(d)} &= \left[\mu_{2} \frac{X^{(d)}Y^{(d)}}{X^{(d)}+k_{2}} - (D+M)Y^{(d)} \right]\, dt\\
	&\qquad + \varepsilon \varsigma_{21} f_{2}\left(R^{(d)},X^{(d)},Y^{(d)}\right)\, dB^{(d)}_{1}(t)+ \varepsilon \varsigma_{22} f_{2}\left(R^{(d)},X^{(d)},Y^{(d)}\right)\, dB^{(d)}_{2}(t),
\end{aligned}
\end{equation}
where D is the dilution rate, $\gamma_{1}$ and $\gamma_{2}$ are the dimensionless yield coefficients for conversion of resource to prey and prey to predator respectively; $\mu_{1}$ and $\mu_{2}$ are the maximum prey and predator specific growth rates; $k_{1}$ and $k_{2}$ are the Monod half saturation constants; $M$ is the rate of loss of predator biomass through endogenous metabolism and $R_0$ the input substrate concentration. The coefficient $\varepsilon$ is a small dimensionless parameter, emphasising that this is a small noise approximation and that 
both noises are of the same order of magnitude. The $\varsigma_{ij}$ determine the 
relative effect of the two noises on the two populations, and the functions $f_{1}$ and $f_{2}$ allow the noise to potentially depend on the resource and biomass concentrations. We explored two specific cases, firstly additive noise, where the functions $f_{i}$ are just constants equal to one, $f_{1} = 1$ and $f_{2} = 1$, which we will refer to as Model (A), and secondly, Model (B), where noise is assumed to enter through the uptake rates to give a multiplicative noise:
\begin{equation}
\begin{aligned}
	f_{1}\left(R^{(d)},X^{(d)},Y^{(d)}\right) &= \mu_{1}\frac{R^{(d)}X^{(d)}}{R^{(d)}+k_{1}}\\
	f_{2}\left(R^{(d)},X^{(d)},Y^{(d)}\right) &=  \mu_{2} \frac{X^{(d)}Y^{(d)}}{X^{(d)}+k_{2}}.
\end{aligned}
\end{equation}
The latter is a more realistic approach since additive noise pre-supposes the spontaneous 
destruction and creation of individuals whereas Model (B) can be mechanistically 
derived from fluctuations in the uptake rates. It is almost certain that there are 
mechanisms where fluctuations in an environmental variable, such as temperature or 
pH \citep{Ratkowsky1982,Ratkowsky1983,Zwietering1991,Rosso1995}, can lead to random fluctuations in uptake rate.

We fixed the biological parameters to values that are typical for bacteria-protozoa systems 
[$\gamma_{1} = 0.4$, $\gamma_{2} = 0.6$, $\mu_{1} = 0.5$, $\mu_{2} = 0.2$, $k_{1} = 8$ and $k_{2} = 9$ - \citep{Nisbet1983}] 
and we chose a value of 0.1 for $M$ as the maximum plausible at half the maximum growth rate. 
In \ref{NML} we also consider the case $M = 0$ but here we focus on this more realistic scenario. 
The input substrate concentration was set at 100 mgL\textsuperscript{-1}, 
substantially lower than the true concentration of nutrients in our substrate (Table \ref{ONE}), 
but these choices for $M$ and $R_{0}$ ensured we were in a regime where coexistence of predator and 
prey occurred at a stable equilibrium up until the dilution rate $D = D_{W} \approx 0.0629$ where 
washout of the predator occurs. It is well known that coexistence is harder to achieve in simple 
predator-prey models than is the case in real complex systems \citep{Nisbet1983}. This value for 
$R_{0}$ allows coexistence without adding extra complexity to our model.

The correlation in this nonlinear model is analytically intractable. However, as we discuss in \ref{NML}, this more realistic non-linear model can be replaced by the linear approximation about its stable equilibrium,
\begin{align*}
	R^{\star} &= \frac{1}{2}\left(k_{1}-R_{0}+\frac{\mu_{1}}{\gamma_{1}} \frac{X^{\star}}{D}\right)
		+ \frac{1}{2}\sqrt{\left(k_{1}-R_{0}+\frac{\mu_{1}}{\gamma_{1}} \frac{X^{\star}}{D}\right)^{2}
		+4R_{0}k_{1}}\\
	X^{\star} &= \frac{k_{2}D}{\mu_{2}-D}\\
	Y^{\star} &= \gamma_{1}\gamma_{2}(R_{0}-R^{\star})-\gamma_{2}X^{\star},
\end{align*}
provided the the noise amplitude is small, an approach that is the continuous-time equivalent of that in \citet{Greenman2005b}.  If we set 
\begin{gather*}
	\ba
	= \begin{pmatrix}
	\frac{\mu_{1}}{\gamma_{1}} \frac{k_{1} k_{2} D}{(D-\mu_{2})(R^{\star}+k_{1})^{2}} - D 	
	& -\frac{\mu_{1}R^{\star}}{\gamma_{1}(R^{\star}+k_{1})} & 0\\
	 -\mu_{1} \frac{k_{1} k_{2} D}{(D-\mu_{2})(R^{\star}+k_{1})^{2}} 
	 &  -\frac{D\left((D-\mu_{1})R^{\star}+D k_{1}\right)}{\mu_{2}(R^{\star}+k_{1})}  
	 & -\frac{D}{\gamma_{2}}\\ 
	0 & \frac{(D-\mu_{2})\left((D-\mu_{1})R^{\star}+D k_{1}\right) }{\mu_{2}(R^{\star}+k_{1})} & 0
	\end{pmatrix},\\
	 \bsigma = \begin{pmatrix}
	 0 & 0 & 0\\  
	 0 & \varsigma_{11} f_{1}\left(R^{\star},X^{\star},Y^{\star}\right)  & \varsigma_{12} f_{1}\left(R^{\star},X^{\star},Y^{\star}\right) \\ 
	 0 & \varsigma_{21} f_{2}\left(R^{\star},X^{\star},Y^{\star}\right) & \varsigma_{22} f_{2}\left(R^{\star},X^{\star},Y^{\star}\right) 
	 \end{pmatrix},
\end{gather*}
and
\begin{align*}
	Z^{(d)}_{1} &= R^{(d)} - R^{\star},\\
	Z^{(d)}_{2} &= X^{(d)} - X^{\star},\\
	Z^{(d)}_{3} &= Y^{(d)} - Y^{\star},\\
\end{align*}
then the linear model
\[	
	dZ^{(1)}_{i}(t) = \left(\sum_{j=1}^{m} a_{ij} Z^{(1)}_{i}(t)\right)\, dt + 
		\sum_{j=1}^{3} \sigma_{ij}\, dB^{(1)}_{j}(t),
\]
$i=1,2,3$, describes the  fluctuations from equilibrium of the populations in the first reactor, whilst
\[	
	dZ^{(2)}_{i}(t) = \left(\sum_{j=1}^{m} a_{ij} Z^{(2)}_{i}(t)\right)\, dt + 
		\sum_{j=1}^{3} \sigma_{ij}\, dB^{(2)}_{j}(t),
\]
$i=1,2,3$, represents those for the second, where the $B^{(d)}_{j}$ are zero mean, unit variance, Brownian motions constructed to be correlated with the corresponding noise in the other community but independent of the other Brownian motions:
\begin{equation}
	\mathbb{E}\left[B^{(1)}_{i}(t),B^{(2)}_{j}(t)\right] = \begin{cases} 
		\rho_{i} & \text{if i=j} \\
		0 & \text{otherwise.} 
	\end{cases}
\end{equation}
Here, the matrices $\ba$, with elements $a_{ij}$, and $\bsigma$, with elements $\sigma_{ij}$, 
give the linear interactions between the population densities and the amplitude of the noise terms.  
 More generally, in \ref{MORAN}, we  consider a system with $m$ species and $n$ noises; 
 writing $\bZ^{(d)}(t) = (Z^{(d)}_{1}(t), \ldots \\ Z^{(d)}_{m}(t))$ for the departure from 
 equilibrium in community $d=1,2$, and  $\ba = (a_{ij})$ and $\bsigma = (\sigma_{ij})$ 
 for the linearised deterministic dynamics and matrix of covariances, respectively, 
then the linearised dynamics are described by an SDE,
\[
	d\bZ^{(d)}(t) = \ba \bZ^{(d)}\, dt + \bsigma d\bB^{(d)}(t).
\]
This SDE can be solved using an integrating factor to give
\begin{equation}\label{IF}
	\bZ^{(d)}(t) = e^{t\ba} \bZ^{(d)}(t) + \int_{0}^{t} e^{(t-s)\ba} \bsigma\, d\bB^{(d)}(t).
\end{equation}

To understand the effect of predator-prey interactions on population synchrony in this linearised 
model requires a multi-species equivalent of the `Moran effect'.   In \ref{MORANM}, we consider 
two identical communities in the general case of $m$ interacting species where each species' 
population density has a growth rate that is a linear function of the other species' 
densities plus a stochastic term. The stochastic element is comprised of $n$ independent 
white noise terms constructed so that each noise term is correlated only with the equivalent 
term in the other community. These different terms represent different environmental factors 
fluctuating independently. The impact of these noise terms on the different species varies and 
is controlled through the coefficient of that noise in the linear dynamics. We are able to 
derive a general expression for the cross-correlation between each species' population density 
in the first community and its population density in the second community as a function of time \eqref{CC}. 
This is a complex expression but for the case when all the noise terms have the same correlation then so
 do all the species \ie the Moran effect applies \eqref{EQRHO}. For a breakdown in synchrony, it is 
 necessary that different noise terms exhibit different amounts of correlation.

We will be interested in the case when $\ba$ is a stable matrix (\textit{i.e.}, all eigenvalues of $\ba$ have real part less than zero).  Then, it is the eigenvalue with smallest real part, that is closest to zero, which is most important in analysing the correlation.   The underlying intuition is simple:  from the solution above, we see that if the stochastic component is being damped in proportion to $e^{t\ba}$, then for any vector $\mathbf{u}$, as $t \to \infty$,  
\begin{equation}
	e^{t\ba} \mathbf{u} \sim e^{\lambda_{\star} t} (\mathbf{u} \cdot \mathbf{f}^{\star}) \mathbf{f}^{\star}, 
\end{equation}
where $\lambda_{\star}$ is the eigenvalue with smallest real part in absolute value, and $\mathbf{f}^{\star}$ is the corresponding eigenvector.  Thus, in the long run, the solution is dominated by the component in the eigenspace corresponding to the eigenvalue $\lambda_{\star}$, and the populations will tend to become synchronised in proportion to the correlation in the noises as projected onto the corresponding eigenvector.  We will refer to this as the `controlling' eigenvector and correspondingly, its eigenvalue as the `controlling' eigenvalue. This terminology allows us to avoid the use of the phrase dominant eigenvector which technically refers to the eigenvector with largest real part, which in an unstable system will dominate the dynamics. Breakdown in synchrony can occur if the interaction coefficients change in response to the control parameters in such a way that the projection of the noise onto the controlling eigenvector increasingly selects the less correlated terms.   In particular, when $\ba$ varies with some control parameter, the dominating eigenvalue and eigenvector will also vary, and with them, the degree of synchrony. 

Using the solution of the SDE, \eqref{IF}, we can explicitly compute the correlation between types \eqref{CORR}, 
for an arbitrary number of species and noises as 
\begin{equation}\label{CORRM}
	\text{corr}(Z^{(1)}_{i}(t),Z^{(2)}_{i}(t)) = 
	\frac{\left(\int_{0}^{t} 
		e^{(t-u)\ba} \bsigma \brho
 \bsigma^{\top} e^{(t-u)\ba^{\top}}\, du\right)_{ii}}
	{\left(\int_{0}^{t} e^{(t-u)\ba} \bsigma \bsigma^{\top} e^{(t-u)\ba^{\top}}\, du\right)_{ii}}
\end{equation}
for
\[
	\brho = \begin{pmatrix} \rho_{1} & & \\ & \ddots & \\ & & \rho_{n} \end{pmatrix}.
\]
From this expressions we see that provided the correlations between the noise components are the 
same for all noise components, \textit{i.e.}, $\brho = \rho I_{n \times n}$, then the 
`Moran effect' always holds: the correlations between both populations equals $\rho$.   

To get a breakdown in synchrony, we require the existence of multiple sources of environmental 
noise with differing degrees of correlation, and that one species is more affected by these noises than the other, 
both plausible assumptions.  To understand better how this occurs, consider the long time limit of the 
cross-correlation in two scenarios.   In the first, we assumed that the noise to which the prey is most 
sensitive (the `prey's noise') is less correlated than that which most affects the predator (the `predator's noise'), 
$\rho_{1} = 0.5$,  $\rho_{2} = 0.9$, and secondly that the predator's noise is less correlated than the prey's noise, 
$\rho_{1} = 0.9$,  $\rho_{2} = 0.5$. In both cases the effect of the noise was taken to be larger for the prey
 population than for the predator ($\sigma_{11} = 0.5$, $\sigma_{12} = 0.3$, $\sigma_{21} = 0.05$, $\sigma_{22} = 0.1$). 
 We then calculated the correlation between the two populations as a function of dilution rates using the method described 
 in the \ref{PPCM}. Essentially, this serves to linearise the equations about the equilibrium so that the 
 interaction matrix $\ba$, and, for Model (B), the noise terms $f_{1}$ and $f_{2}$, become a function of the 
 equilibrium concentrations (which vary with the dilution rate $D$). 

The results are shown in Figures 3 and 4 for Model (A) and Model (B) respectively. Clearly, it is only 
in the first case with the predator's noise more correlated than the prey's noise that we can 
reproduce both the synchrony breakdown and the observation that the predator densities were more correlated. 
In this instance, for both models, the prey and predator correlations converge on that of the predator's noise at 
zero dilution rate. In the \ref{appapp}, we explain that all species having the same correlation is a general 
feature of points at which the real part of the controlling eigenvalue goes to zero, and in this case, it is the 
`predator's noise'  that dominates the system. In both models, as the dilution rate is increased, 
there is a loss of synchrony as the influence of the prey's noise becomes more significant, but this 
changes again with further increase in the dilution rate. In fact with Model (A)  - Fig. 3a - as we 
approach washout, the controlling eigenvalue approaches zero again, and both correlation in both 
species' density converges onto the correlation of the predator's noise. The third eigenvalue also 
goes to zero in Model (B) -- the second and third eigenvalues are shown as a function of dilution 
rate for this model in Figure 5.  However, in this instance, we do not see the population 
correlations converge, essentially because the noise associated with the predator goes to 
zero as the predator experiences washout from the system.  The predator is then no longer 
able to influence the prey dynamics, which is dominated by its own relatively low noise 
correlation. We prefer the more realistic form of noise in Model (B), and the 
corresponding interpretation of the dynamics. It also fits the observations better, 
with the breakdown in prey synchrony in Fig. 4a corresponding qualitatively to what we observe. 
The model also fits the observed changes in predator correlation at least until the correlation 
starts to increase again. However, this increase is not substantial until dilution rates are 
larger than those realised experimentally, roughly $D = 0.05$, corresponding to $d = 0.30$, 
accounting for the factor of six necessary to transform time scales. 

\section{Discussion} 

We observed the novel phenomenon of a synchrony breakdown between replicate reactors as dilution rate increases.  This occurred for both bacterial and protozoal dynamics. Moreover, we were able to use mathematical models to propose a hypothesis to explain this phenomenon: the breakdown in synchrony can be explained in terms of the controlling eigenvalues and the projection onto the corresponding eigenvector of the linearised population dynamics. This assumes a predator (protozoa) population that primarily experiences highly correlated environmental noise and a prey (bacteria) population that is primarily affected by a less correlated noise. The changes in both bacterial and protozoa synchronies were consistent with this hypothesis, but only if both populations experienced multiplicative rather than additive noise, which is almost certainly the case. We have also demonstrated the utility of simple batch reactors as a tractable means of exploring synchrony in multi-species communities.  The use of a complex, `real' microbial community to investigate synchrony builds on previous examples of population studies in artificially constructed, simple, 2-3 species experiments \citep{Fontaine2005, Becks2005, Vasseur2009,Fox2011}.  

If our hypothesis is correct, then this is the first experimental example of trophic interactions mediating the transfer of noise in communities, something previously predicted but not observed \citep{Greenman2005b}. In fact, given that the bacterial communities in these reactors will comprise many species, each with somewhat different demographics, then to have even observed the `Moran effect' at this aggregate level is surprising. To observe synchrony breaking down in a coherent way with operating conditions is doubly so, and a dramatic confirmation of the value of multi-species reactors for testing ecological theory \citep{Benton2007}. 

This result provides an elegant example of how theory and experiment can inform each other in ecology.  In addition, our hypothesis raises a number of important considerations for the understanding of synchrony in natural multi-species communities: the observation that different trophic groups can respond to the same noisy environment differently, to the extent that they experience effectively independent noise components with different degrees of correlation, should not be overlooked. We predict that the predators, in this example protozoa, primarily experience noise with a higher degree of intrinsic correlation.  We do not know the reason for this, but it may reflect their potentially longer lifespan, which cannot be incorporated into the non-structured dynamical models used here.  

To put these results in a more general context, we observe that dilution rate is the inverse of retention time, in aquatic systems, therefore we might expect similar patterns of synchrony breakdown in microbial populations between systems of small lakes as retention time decreases. Synchrony has been previously observed between lake microbes \citep{kent2007} so this prediction could be directly tested. For terrestrial systems, dilution rate lacks an exact analogue, but whenever we have a system where different trophic groups respond to the different components of noise in the environment differently, then we could have synchrony changes as environmental factors cause changes in the importance of those sources of noise to the overall community dynamics. For instance, in a simple predator-prey system, increased input of external resources should cause a shift from prey to predator dominated dynamics and possibly an increase in synchrony amongst isolated populations. 

Further experiments would help confirm the hypotheses raised by this study, for example, increasing  reactor dilution rates until predator washout occurs. If this was accompanied by a continued decrease in microbial biomass correlation, but an increase in protozoa correlation, it would be a dramatic confirmation of our predictions. Other operating conditions could also be altered, such as the resource concentration, and the  changes in synchrony then observed could be compared to our mathematical predictions. Alternatively, it would be intriguing to fully resolve the microbial biomass component into the different bacterial and protozoan taxa present, in which case we would expect that a distribution of correlations for the different species would be observed. Such detailed experimental work would provide data sets suitable for expanding the mathematical model to multiple species within the same trophic level or functional group of the community, defined by the consumption of a particular nutrient, or by a particular ecosystem function. In general, we believe that experimental microbial communities coupled to the mathematical framework presented here, will prove to be a powerful paradigm for the study of synchrony in multi-species communities.

\section*{Acknowledgements}

We wish to thank Russell Davenport and Ozge Eyice for assistance with the experimental work and 
Jan Lindstrom for helpful comments on this manuscript. 
CQ was supported by an Engineering and Physical Sciences Research Council Career Acceleration Fellowship (EP/H003851/1) and 
an MRC fellowship as part of the CLIMB consortium (MR/L015080/1).  
Part of the work presented here was done whilst TLP was supported by a Fondation Sciences Math\'ematiques de
 Paris Postdoctoral Fellowship. The experimental work was conducted under an EPSRC Platform Grant (GR/S59543/01).

\bibliographystyle{abbrvnat} 
\bibliography{Synchrony}

\clearpage

\begin{table}

\begin{tabular}{p{.08\textwidth}p{.16\textwidth}p{.16\textwidth}p{.1\textwidth}p{.18\textwidth}p{.18\textwidth}}
Dilution fraction $d$ & Temp. (\degree C) Rep. A (mean, std. dev.) & Temp. (\degree C) Rep. B (mean, std. dev.) & Temp. Cross-correlation 
	& COD conc. (mgL\textsuperscript{-1}) (mean, std. dev.) & NH4-N conc. (mgL\textsuperscript{-1}) (mean, std. dev.)\\
\hline 
0.025 & 13.1, 0.38 & 13.6, 0.37 & 0.88 & 575.2, 45.2 & 33.1, 9.5 \\
0.035 & 13.6, 0.24 & 13.8, 0.17 & 0.69 & 602.1, 35.6 & 36.1, 1.5 \\
0.0833 & 14.6, 0.24 &14.3, 0.25 & 0.74 & as $d = 0.035$  & as $d = 0.035$\\
0.25 & 14.3, 0.52 & 14.0, 0.51 & 0.98 & as $d = 0.025$  & as $d = 0.025$\\
\hline
\end{tabular}
\caption{Summary of temperatures at the four dilution rates for both replicates A and B together with the un-lagged cross-correlation calculated using the ccf function of R. The mean and std. dev of the COD (chemical oxygen demand) and NH4-N (nitrogen concentration) in the influent substrate.} 
\label{ONE}
\end{table}

\clearpage

\begin{figure}
\begin{center}
	\includegraphics[width=.6\linewidth]{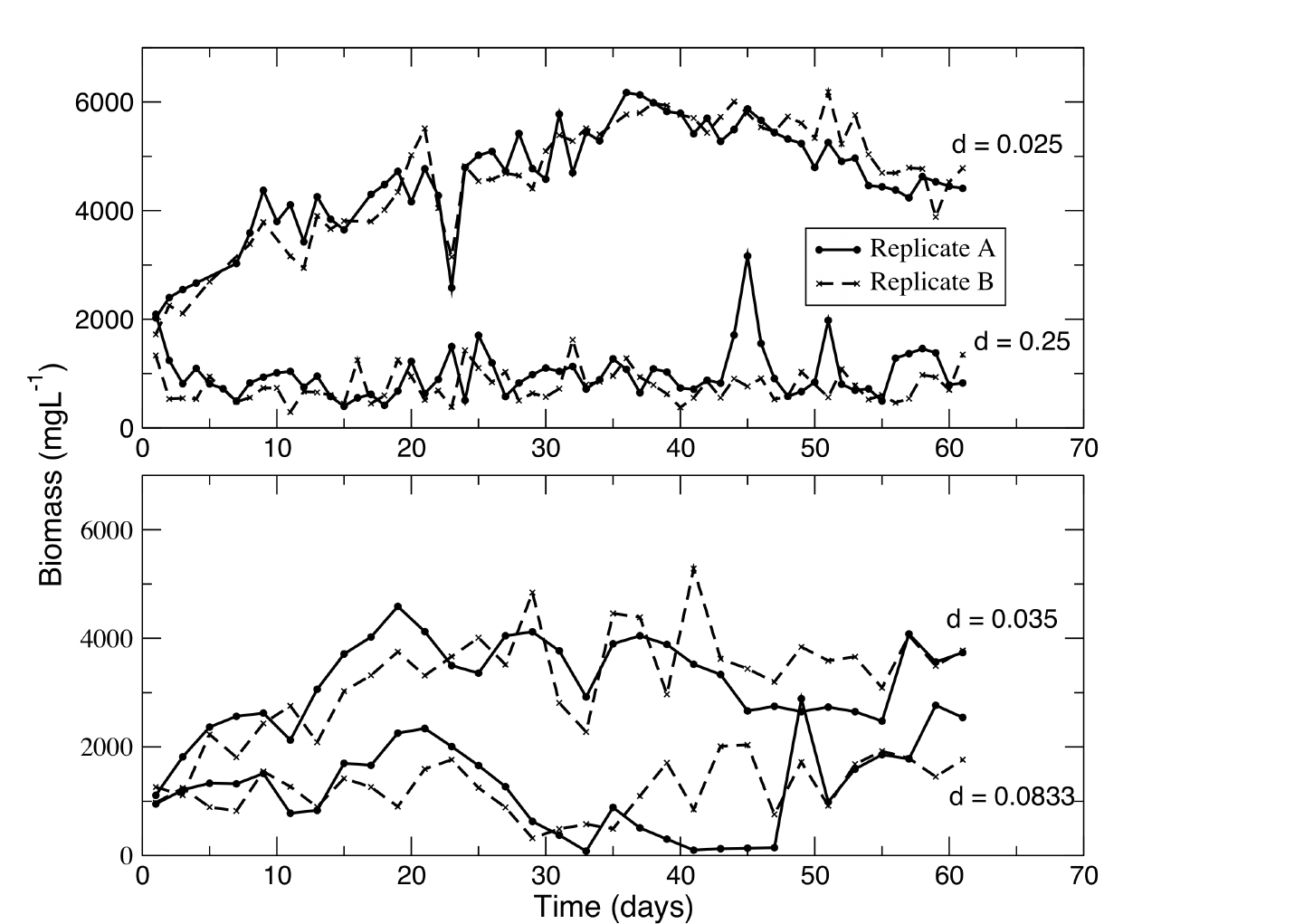}
\end{center}
\caption{Microbial biomass as a function of time for both replicates at the four different dilution rates.}
\end{figure}

\begin{figure}
\begin{center}
	\includegraphics[width=.6\linewidth]{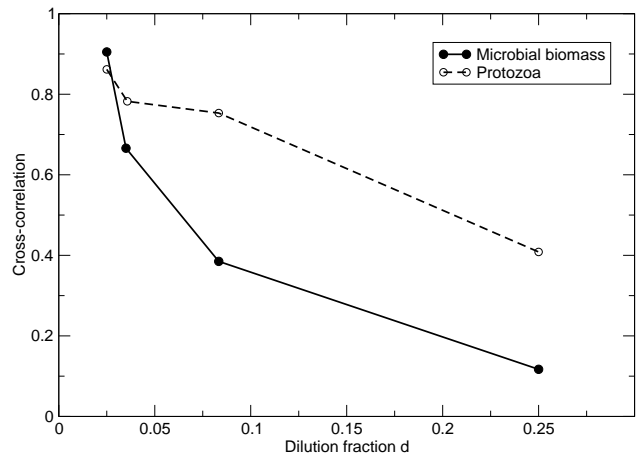}
\end{center}
\caption{The microbial biomass and protozoa cross-correlations (computed using R ccf function at zero lag) between replicate reactors as a function of the dilution fraction.} 
\end{figure}

\clearpage

\begin{figure}
\begin{center}
\mbox{
	\subfigure[]{\includegraphics[width=.4\textwidth]{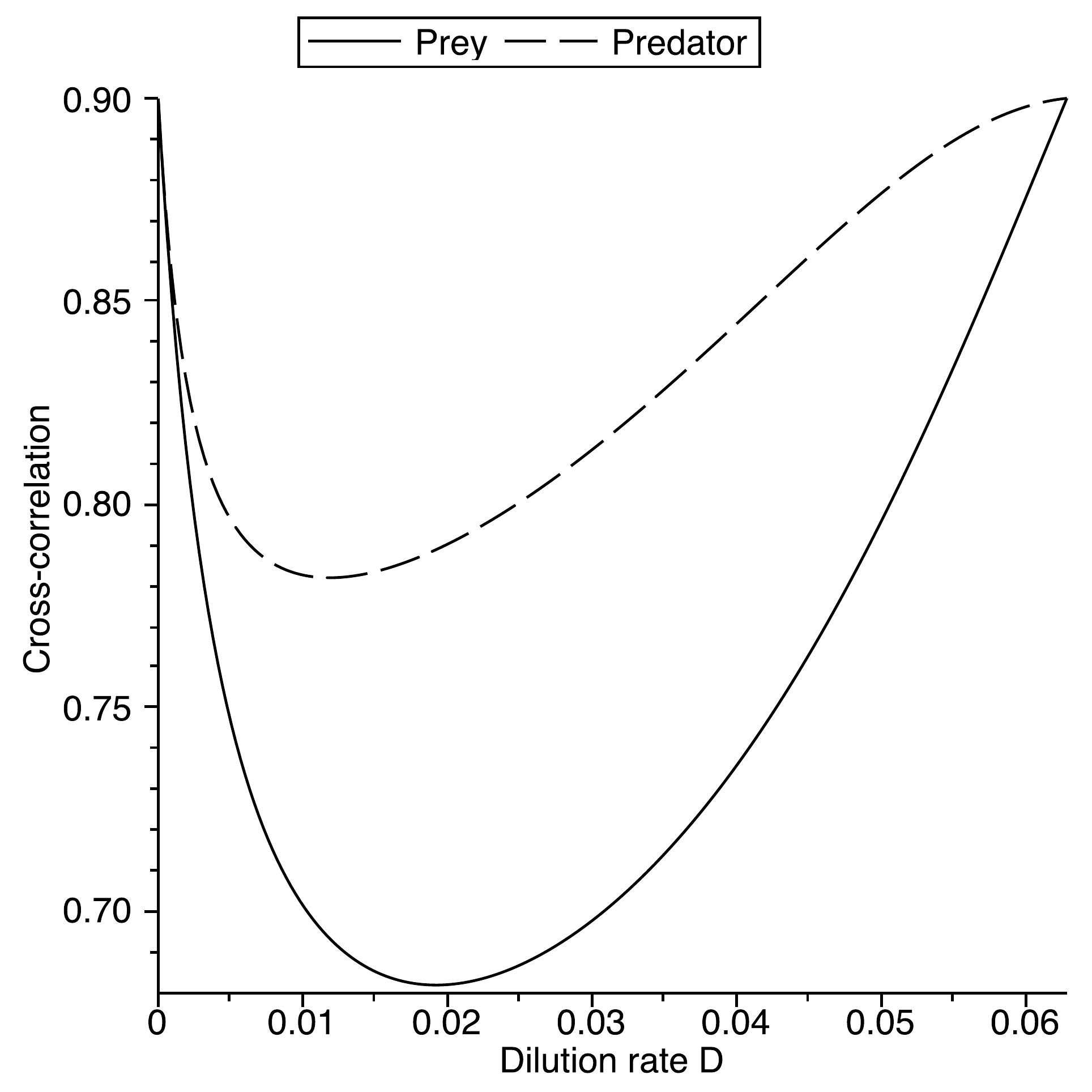}}
	\subfigure[]{\includegraphics[width=.4\textwidth]{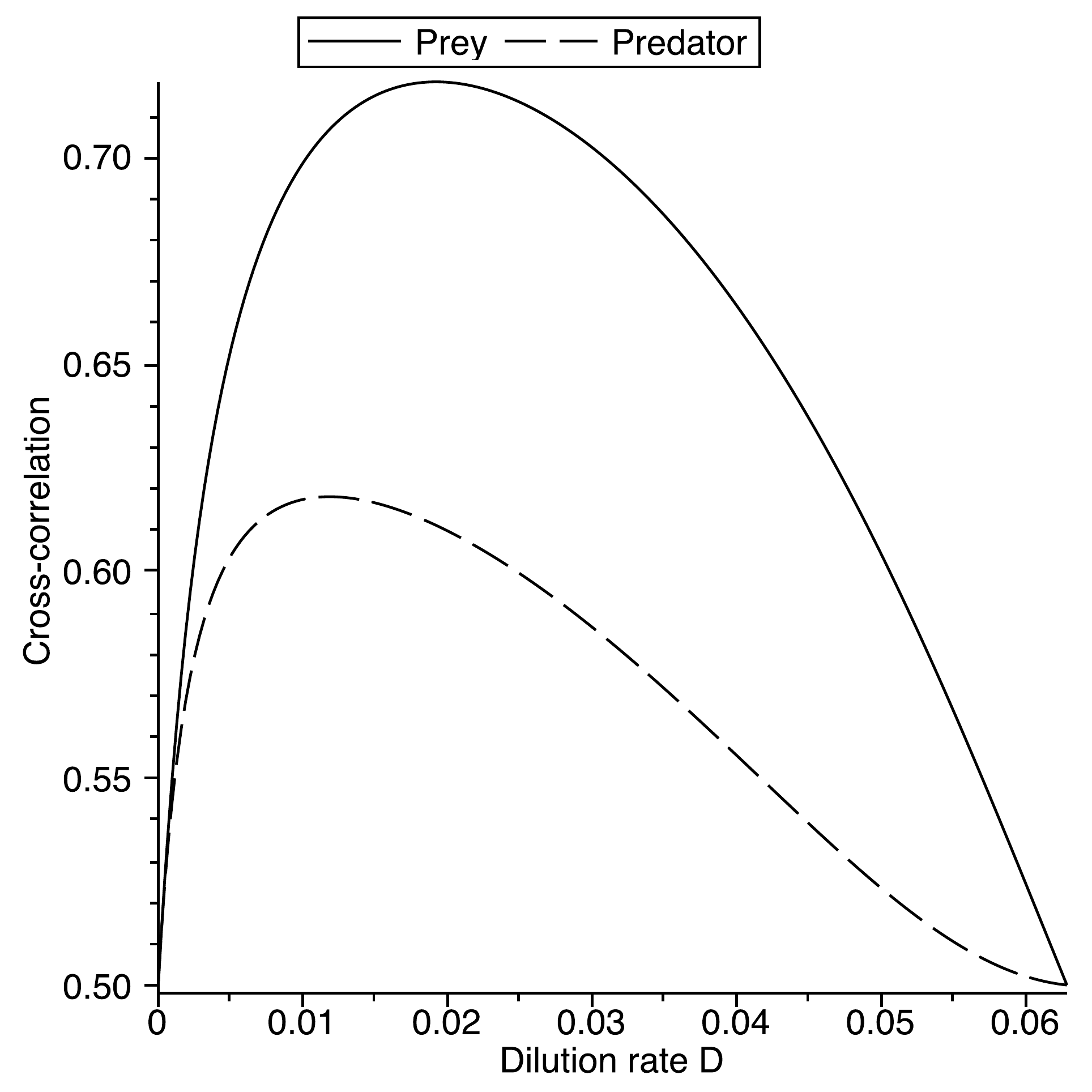}}}
\end{center}
\caption{Cross-correlation in predator and prey populations for model (A - additive noise) with $M = 0.1$, $\sigma_{11} = 0.5$, $\sigma_{12} = 0.3$, $\sigma_{21} = 0.05$, $\sigma_{22} = 0.1$,  (a) $\rho_{1} = 0.5$, $\rho_{2} = 0.9$, and (b) $\rho_{1} = 0.9$, $\rho_{2} = 0.5$.}
\end{figure}

\begin{figure}
\begin{center}
\mbox{
	\subfigure[]{\includegraphics[width=.4\textwidth]{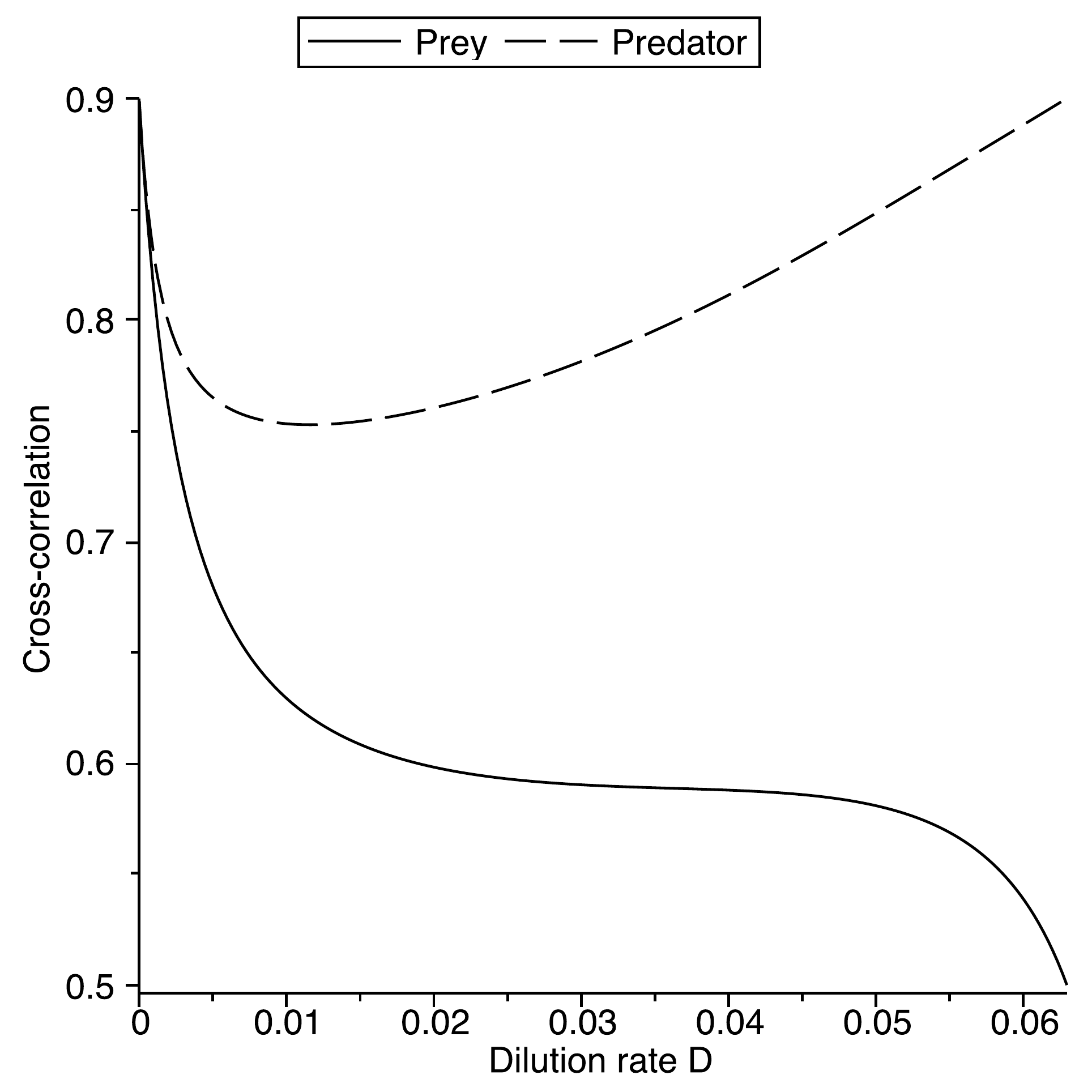}}
	\subfigure[]{\includegraphics[width=.4\textwidth]{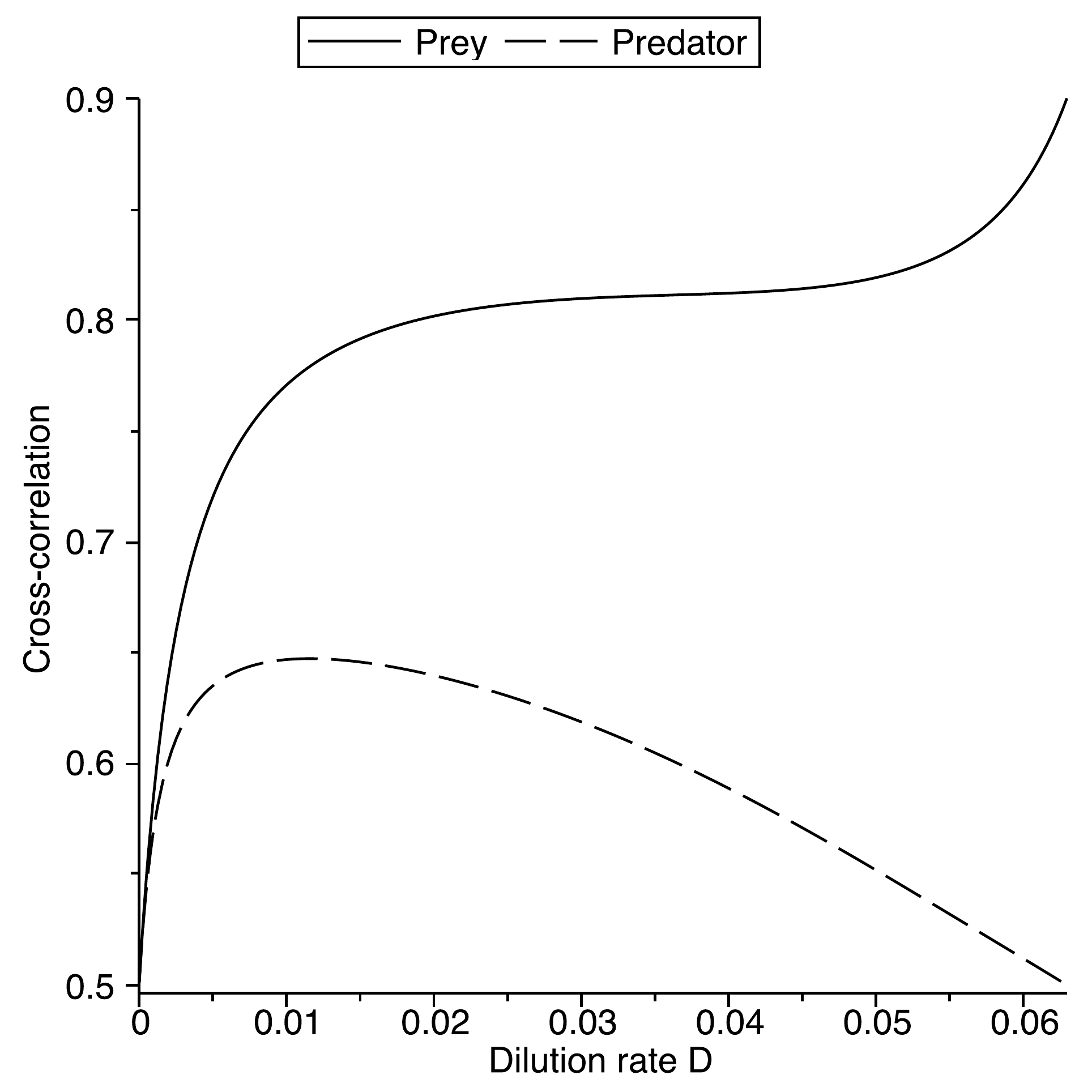}}}
\end{center}
\caption{Cross-correlation in predator and prey populations for model (B - multiplicative noise) with $M = 0.1$, $\sigma_{11} = 0.5$, $\sigma_{12} = 0.3$, $\sigma_{21} = 0.05$, $\sigma_{22} = 0.1$, (a) $\rho_{1} = 0.5$, $\rho_{2} = 0.9$, and (b) $\rho_{1} = 0.9$, $\rho_{2} = 0.5$.}
\end{figure}
\begin{figure}
\begin{center}
	\includegraphics[width=.6\linewidth]{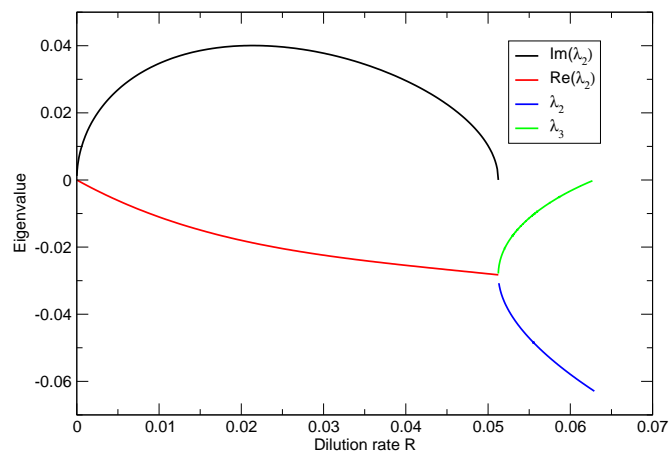}
\end{center}
\caption{Eigenvalues corresponding to Figure 4A) corresponding to model (B - multiplicative noise) with $M = 0.1$. The first eigenvalue $\lambda_{1}$ is not shown as it is always large and negative and hence does not impact the synchrony.}
\end{figure}

\clearpage

\appendix
\renewcommand*{\thesection}{\appendixname~\Alph{section}}

\section{The Moran Effect in One or More Dimensions}\label{MORAN}

Moran's result \citep{Moran1953} was originally demonstrated in a 
discrete-time model with distinct generations, but the argument applies 
equally well to a population growing in continuous time; we give a
 brief proof below.  

\subsection{One Dimension}

Suppose the number of individuals in population $i$ ($i=1,2$) satisfies
\begin{equation}\label{OU}
	dX_i(t) = a X_i(t)\, dt + \sigma\, dB_i(t)
\end{equation}
for correlated Brownian motions $B_1(t)$ and $B_2(t)$:
\[
	\E{B_1(s)B_2(t)} = \rho (s \wedge t) \qquad (\rho \leq 1)
\]
Now \eqref{OU} is the stochastic differential equation (SDE) for an Ornstein-Uhlenbeck process, and may be solved explicitly:
\[
	X_i(t) = e^{at} X_i(0) + \int_{0}^{t}  e^{a(t-s)}\sigma\, dB_i(s).
\]
(\textit{N.B.}, we use the It\=o SDE throughout).  Thus $X_i(t)$ has mean $e^{at} X_i(0)$ and variance 
\[
	\int_{0}^{t} e^{2a(t-s)}\sigma^{2}\, ds =  \frac{e^{2at}-1}{2a}\sigma^{2},
\]
whilst
\[
	\text{Cov}(X_1(s),X_2(t)) = \int_{0}^{s \wedge t}  e^{2a(t-s)} \rho \sigma^{2}\, ds 
\]
so that 
\begin{align*}
	\text{corr}(X_1(t),X_2(t)) 
	&\defn \frac{\text{Cov}(X_1(t),X_2(t))}{\sqrt{\text{Var}(X_1(t))}\sqrt{\text{Var}(X_2(t))}}\\
	&=\rho, 
\end{align*}
from which we conclude that synchronous noises lead to synchronous population dynamics.

\subsection{$m$ Dimensions}
\label{MORANM}
Let $B_{i}(t)$, $i=1,\ldots,2n$ be independent Brownian motions.  We use these to construct pairs of correlated Brownian motions: for $i=1,\ldots,n$, let 
\begin{gather*}
	B^{(1)}_{i}(t) = B_{i}\\
	B^{(2)}_{i}(t) = \rho_{i} B_{i}(t) + (1-\rho_{i}^{2})^{\frac{1}{2}} B_{n+i}(t),
\end{gather*}
so that 
\[
	\E{B^{(1)}_{i}(t),B^{(2)}_{j}(s)}
	= \begin{cases} 
		\rho_{i}t & \text{if i=j} \\
		0 & \text{otherwise.} 
	\end{cases}
\]
We now consider a system of two pairs of coupled linear SDEs, representing two populations of 
$m$ interacting species with identical deterministic dynamics and experiencing correlated noise: 
\[	
	dX_{1i}(t) = \left(\sum_{j=1}^{m} a_{ij} X_{1i}(t)\right)\, dt + 
		\sum_{j=1}^{n} \sigma_{ij}\, dB^{(1)}_{j}(t)
\]
describe the populations in the first patch, whilst
\[	
	dX_{2i}(t) = \left(\sum_{j=1}^{m} a_{ij} X_{2i}(t)\right)\, dt + 
		\sum_{j=1}^{n} \sigma_{ij}\, dB^{(2)}_{j}(t)
\]
represent those in the second.  This system can then be expressed compactly using matrix 
equations.  Let $\ba = (a_{ij})$, $\bsigma = (\sigma_{ij})$, and 
\begin{gather*}
	\bX_{i}(t) = \mat{X_{i1}(t) \\ \vdots \\ X_{im}(t)}, \quad \text{and} \quad
	\bX(t) = \mat{\bX_{1}(t) \\  \bX_{2}(t)},\\
	\bB(t) = \mat{B_{1}(t)\\ \vdots \\B_{n}(t)},\\
\intertext{and}
	\brho
 = \mat{\rho_{1} & & \\ & \ddots & \\ & & \rho_{n}} \quad \text{and} \quad
	\tilde{\brho} = 
	\mat{(1-\rho_{1}^{2})^{\frac{1}{2}} & & \\ & \ddots & \\ & & (1-\rho_{2}^{n})^{\frac{1}{2}}}.
\end{gather*}
Then, 
\[
	d\bX(t) = \mat{\ba & 0 \\ 0 & \ba} \bX(t)\, dt
	+ \mat{\bsigma & 0 \\ \bsigma \brho & \bsigma \tilde{\brho}}\, d\bB(t) 
\]

This linear system admits an exact solution, 
\begin{align}\label{ES}
	\bX(t) &= \mat{e^{t\ba} & 0 \\ 0 & e^{t\ba}} \bX(0)
		+ \int_{0}^{t} \mat{e^{(t-u)\ba} & 0 \\ 0 & e^{(t-u)\ba}} 
		\mat{\bsigma & 0 \\ \bsigma \brho
 & \bsigma \tilde{\brho}}\, d\bB(u)\\
	&= \mat{e^{t\ba} & 0 \\ 0 & e^{t\ba}} \bX(0)
		+ \int_{0}^{t} \mat{e^{(t-u)\ba} \bsigma & 0 \\
		e^{(t-u)\ba} \bsigma \brho
 & e^{(t-u)\ba} \bsigma \tilde{\brho}}\, d\bB(u)
\end{align}
so that if $\bX(0) = X_{0}$ is fixed, 
\[
	\E{\bX(t)} =  \mat{e^{t\ba} & 0 \\ 0 & e^{t\ba}} \bX_{0},
\]
whilst the variance-covariance matrix, 
\[
	\text{Var}(\bX(t)) = \E{\left(\bX(t)-\E{\bX(t)}\right)\left(\bX(t)-\E{\bX(t)}\right)^{\top}}
\]
is
\begin{multline*}
	\int_{0}^{t} \mat{e^{(t-u)\ba} \bsigma & 0 \\
		e^{(t-u)\ba} \bsigma \brho
 & e^{(t-u)\ba} \bsigma \tilde{\brho}}
		\mat{e^{(t-u)\ba} \bsigma & 0 \\
		e^{(t-u)\ba} \bsigma \brho
 & e^{(t-u)\ba} \bsigma \tilde{\brho}}^{\top}\, ds\\
	= \int_{0}^{t} \mat{e^{(t-u)\ba} \bsigma \bsigma^{\top} e^{(t-u)\ba^{\top}}
		& e^{(t-u)\ba} \bsigma \brho
^{\top} \bsigma^{\top} e^{(t-u)\ba^{\top}}  \\
		e^{(t-u)\ba} \bsigma \brho
 \bsigma^{\top} e^{(t-u)\ba^{\top}} 
		& e^{(t-u)\ba} \bsigma (\brho
 \brho
^{\top} + \tilde{\brho} \tilde{\brho}^{\top}) \bsigma^{\top}e^{(t-u)\ba^{\top}}}
		\, ds\\
	= \int_{0}^{t} \mat{e^{(t-u)\ba} \bsigma \bsigma^{\top} e^{(t-u)\ba^{\top}}
		& e^{(t-u)\ba} \bsigma \brho
^{\top} \bsigma^{\top} e^{(t-u)\ba^{\top}}  \\
		e^{(t-u)\ba} \bsigma \brho
 \bsigma^{\top} e^{(t-u)\ba^{\top}} 
		& e^{(t-u)\ba} \bsigma \bsigma^{\top}e^{(t-u)\ba^{\top}}}\, ds.
\end{multline*}
Thus, the cross-covariance between $\bX_{1}(t)$ and $\bX_{2}(t)$ is
\begin{equation}\label{COV}
	\text{Cov}(\bX_{1}(t),\bX_{2}(t)) 
		= \int_{0}^{t} e^{(t-u)\ba} \bsigma \brho
 \bsigma^{\top} e^{(t-u)\ba^{\top}}\, du
\end{equation}
whilst for both, 
\begin{equation}\label{VAR}
	\text{Var}(\bX_{i}(t)) 
	= \int_{0}^{t} e^{(t-u)\ba} \bsigma \bsigma^{\top} e^{(t-u)\ba^{\top}}\, du
\end{equation}
so that the cross-correlation is
\begin{equation}
\label{CC}
	\text{corr}(X_{1i}(t),X_{2j}(t)) = 
	\frac{\left(\int_{0}^{t} 
		e^{(t-u)\ba} \bsigma \brho
 \bsigma^{\top} e^{(t-u)\ba^{\top}}\, du\right)_{ij}}
	{\left(\int_{0}^{t} e^{(t-u)\ba} \bsigma \bsigma^{\top} e^{(t-u)\ba^{\top}}\, du\right)_{ii}
		^{\frac{1}{2}}
	\left(\int_{0}^{t} e^{(t-u)\ba} \bsigma \bsigma^{\top} e^{(t-u)\ba^{\top}}\, du\right)_{jj}
		^{\frac{1}{2}}}.
\end{equation}
In particular, we observe that when 
\[
	\brho = \rho I_{n \times n},
\]
then the Moran effect holds \text{i.e.}
\begin{equation}\label{EQRHO}
	\text{corr}(X_{1i}(t),X_{2i}(t)) = \rho.
\end{equation}

\subsection{Short and Long Term Behaviour}\label{ShortLong}

Using \eqref{COV}, \eqref{VAR} and \eqref{CORR}, we can derive asymptotic expressions for the correlation as $t \to 0$ and as $t \to \infty$.  For simplicity, we will confine ourselves to the cases of interest, 
\begin{equation}\label{CORR}
	\text{corr}(X_{1i}(t),X_{2i}(t)) = 
	\frac{\left(\int_{0}^{t} 
		e^{(t-u)\ba} \bsigma \brho
 \bsigma^{\top} e^{(t-u)\ba^{\top}}\, du\right)_{ii}}
	{\left(\int_{0}^{t} e^{(t-u)\ba} \bsigma \bsigma^{\top} e^{(t-u)\ba^{\top}}\, du\right)_{ii}}
\end{equation}

Examining \eqref{CORR}, we see that both the numerator and denominator vanish as $t \to 0$; to proceed, we use l'H\^opital's Rule, observing that 
\begin{multline}\label{DCOV}
	\frac{d}{dt} \int_{0}^{t} e^{(t-u)\ba} \bsigma \brho
 \bsigma^{\top} e^{(t-u)\ba^{\top}}\, du
	 = \bsigma \brho
 \bsigma^{\top} + 
	 \ba  \int_{0}^{t} e^{(t-u)\ba} \bsigma \brho
 \bsigma^{\top} e^{(t-u)\ba^{\top}}\, du\\ 
	 +  \int_{0}^{t} e^{(t-u)\ba} \bsigma \brho
 \bsigma^{\top} e^{(t-u)\ba^{\top}}\, du\, \ba^{\top}
\end{multline}
and similarly for the variance.  Thus, 
\begin{align*}
	\lim_{t \to 0} \text{corr}(X_{1i}(t),X_{2i}(t)) 
	= \frac{\left(\bsigma \brho
 \bsigma^{\top}\right)_{ii}}{\left(\bsigma \bsigma^{\top}\right)_{ii}}
	= \frac{\sigma_{i1}^{2}}{\sum_{k=1}^{n} \sigma_{ik}^{2}}\rho_{1} + \cdots
		+ \frac{\sigma_{in}^{2}}{\sum_{k=1}^{N} \sigma_{ik}^{2}}\rho_{n} 
\end{align*}

To obtain the limiting behaviour as $t \to \infty$, we observe that \eqref{DCOV} gives us a differential equation for $\text{Cov}(\bX_{1}(t),\bX_{2}(t))$ as a function of $t$, for which 
\[
	\bc \defn \lim_{t \to \infty} \int_{0}^{t} e^{(t-u)\ba} \bsigma \brho
 \bsigma^{\top} e^{(t-u) \ba^{\top}}\, du
\]
is a rest point.  Thus, from \eqref{DCOV}, we have
\begin{equation}\label{MATEQ}
	\ba \bc + \bc \ba^{\top} = -  \bsigma \brho \bsigma^{\top},
\end{equation}	
and similarly for asymptotic variance, $\bv = \lim_{t \to \infty} \text{Var}(\bX_{i}(t))$,
\[
		\ba \bv + \bv \ba^{\top} = -  \bsigma \bsigma^{\top},
\]
These matrix equation may be transformed into a system of linear equations and solved by identifying the space of $m \times m$ matrices with real entries, $M_{m \times m}(\R)$, with the vector space $\R^{m^{2}}$ via the basis $\{\be^{(i,j)}\}_{i,j=1}^{m}$, where $\be^{(i,j)}$  is the $m \times m$ matrix with $ij$\textsuperscript{th} entry equal to one, and all other entries equal to zero (see \citet{Horn+Johnson91}, \S 4.4). Adopting a lexicographical ordering on this basis,  we have the vectorization of the matrix, 
 \[
	\text{vec}: M_{m \times m}(\R) \to \R^{m^{2}}
\]
where 
\[
	\text{vec}(\bx)_{(i-1)m+j} = x_{ij}.
\] 
We may then rewrite \eqref{MATEQ} as 
\begin{equation}\label{TENEQ}
	(\ba \otimes I_{m \times m} + I_{m \times m} \otimes \ba) C 
	= -\text{vec}(\bsigma \brho \bsigma^{\top})
	= -\sum_{k=1}^{n} \rho_{k} \Sigma^{(k)},
\end{equation}
where  $\otimes$ denotes the Kronecker product of matrices, 
\begin{equation}\label{DEFCSIGMA}
	C = \text{vec}(\bc) = \mat{c_{11}\\ \vdots \\ c_{1m}\\ \vdots \\ c_{m1}\\ \vdots\\ c_{mm}}, 
	\quad \text{and} \quad
	\Sigma^{(k)} \defn
	\mat{\sigma_{1k}^{2} \\ \sigma_{1k}\sigma_{2k} \\ \vdots \\ \sigma_{1k}\sigma_{mk} \\
	\vdots \\ \sigma_{mk}\sigma_{1k} \\ \vdots \\ \sigma_{mk}\sigma_{(m-1)k} \\ \sigma_{mk}^{2}}.
\end{equation}
Substituting $\rho_{1} = \cdots = \rho_{m} =1 $ gives an equation for the asymptotic variance. \textit{e.g.} for $m = n = 2$, we have
\[
	\mat{2 a_{11} & a_{12} & a_{12} & 0\\ a_{21} & a_{11} + a_{22} & 0 & a_{12}\\
		a_{21} & 0 & a_{11} + a_{22} & a_{12}\\ 0 & a_{21} & a_{21} & 2 a_{22}}
	\mat{c_{11}\\ c_{12}\\ c_{21}\\ c_{22}} = 
	\rho_{1} \mat{\sigma_{11}^{2} \\ \sigma_{11}\sigma_{21} \\ \sigma_{21}\sigma_{11} \\ 
	\sigma_{21}^{2}} + \rho_{2}  \mat{\sigma_{12}^{2} \\ \sigma_{12}\sigma_{22} \\ 
	\sigma_{22}\sigma_{12} \\ \sigma_{22}^{2}}.
\]
Solving these by Gaussian elimination or Cramer's rule, we then have
\[
	\lim_{t \to \infty} \text{corr}(X_{1i}(t),X_{2i}(t)) = \frac{c_{ii}}{v_{ii}}.
\]
For $n=m=2$, we thus obtain an explicit expression for the asymptotic correlation,
\begin{multline*}
	\lim_{t \to \infty}  \text{corr}(X_{11}(t),X_{21}(t))  = 
	\frac{\det(\ba) \sigma_{11}^{2} + (a_{22} \sigma_{11} - a_{12} \sigma_{21})^{2}}
	{\det(\ba) (\sigma_{11}^{2}+\sigma_{12}^{2} ) + (a_{22} \sigma_{11} - a_{12} \sigma_{21})^{2}
	+ (a_{22} \sigma_{12} - a_{12} \sigma_{22})^{2}} \rho_{1}\\
	+\frac{\det(\ba) \sigma_{12}^{2} + (a_{22} \sigma_{12} - a_{12} \sigma_{22})^{2}}
	{\det(\ba) (\sigma_{11}^{2}+\sigma_{12}^{2} ) + (a_{22} \sigma_{11} - a_{12} \sigma_{21})^{2}
	+ (a_{22} \sigma_{12} - a_{12} \sigma_{22})^{2}} \rho_{2}
\end{multline*}
and
\begin{multline*}
	\lim_{t \to \infty}  \text{corr}(X_{12}(t),X_{22}(t)) = 
	\frac{\det(\ba) \sigma_{21}^{2} + (a_{11} \sigma_{21} - a_{21} \sigma_{11})^{2}}
	{\det(\ba) (\sigma_{21}^{2}+\sigma_{22}^{2} ) + (a_{11} \sigma_{21} - a_{21} \sigma_{11})^{2}
	+ (a_{11} \sigma_{22} - a_{21} \sigma_{12})^{2}} \rho_{1}\\
	+\frac{\det(\ba) \sigma_{22}^{2} + (a_{11} \sigma_{22} - a_{21} \sigma_{12})^{2}}
	{\det(\ba) (\sigma_{21}^{2}+\sigma_{22}^{2} ) + (a_{11} \sigma_{21} - a_{21} \sigma_{11})^{2}
	+ (a_{11} \sigma_{22} - a_{21} \sigma_{12})^{2}} \rho_{2}.
\end{multline*}
Unfortunately, we were  unable to find similarly tidy expressions for $m,n \geq 2$.

\section{Nonlinear Models and Linearisation}\label{NML}

As a model of populations subject to noise, however, \eqref{OU} leaves much to be desired; the 
deterministic component, whilst it may be obtained from a linear birth-death process in the limit of 
large initial numbers, leads to exponential growth or decay, whilst the additive noise lacks an 
obvious mechanistic interpretation other than spontaneous creation or loss of individuals from and 
to the ether.

Several studies have suggested that \eqref{OU} may be still be recovered as a linear 
approximation to a nonlinear process.  Indeed, it is shown in \citet{Freidlin+Wentzell98} 
that if $\varepsilon > 0$ is sufficiently small, then the nonlinear process
\[
	d\bX(t) = \bF(\bX(t))\, dt + \varepsilon\bSigma(\bX(t))\, d\bB(t)
\]
may be approximated to $\BigO{\varepsilon^2}$ on a fixed time interval $[0,T]$ by 
$\bX^{(0)}(t) + \varepsilon \bX^{(1)}(t)$, where $\bX^{(0)}(t)$ is the solution to the deterministic ODE
\begin{equation}\label{NLDS}
	\frac{d}{dt} \bX^{(0)}(t) = \bF(\bX^{(0)}(t)),
\end{equation}
whilst $\bX^{(1)}(t)$ satisfies a linear (Ornstein-Uhlenbeck) SDE:
\[
	d\bX^{(1)}(t) =  (\bD\bF)(\bX^{(0)}(t)) \bX^{(1)}(t)\, dt + \bSigma(\bX^{(0)}(t))\, d\bB(t).
\]
We then have
\begin{equation}\label{LINEARERROR}
	\sup_{0 \leq t \leq T} \left(\E{\abs{\bX(t) - \bX^{(0)}(t) - \varepsilon \bX^{(1)}(t)}^{2}}\right)
		^{\frac{1}{2}} \leq C \varepsilon^{2}
\end{equation}
for some constant $C$ depending on $T$.

In particular, taking $\bX^{(0)}(0)$ to be an asymptotically stable fixed point  $\bx^{\star}$ of \eqref{NLDS},  and setting
\[
	\ba = (\bD\bF)(\bx^{\star}) \quad \text{and} \quad \bsigma = \bSigma(\bx^{\star}),
\]
we recover the linear models of the previous section, and thus are able to apply our results on 
synchrony after a transient period during which the system approaches equilibrium. 

\subsection{The Predator-Prey Chemostat Model}\label{PPCM}

In this section, we will apply the results of the previous section to the predator-prey chemostat dynamics discussed in the main text.  To simplify the analysis, we will assume a continuous flow-through of fresh medium at rate $D$.

We will consider two models of a pair of chemostats, $d=1,2$ experiencing correlated noise:
\begin{itemize}
\item[(A)] Additive noise:
\begin{align*}
	dR^{(d)} &=
	\left[(R_{0}-R^{(d)})D - \frac{\mu_{1}}{\gamma_{1}} \frac{R^{(d)}X^{(d)}}{R^{(d)}+k_{1}}\right]\, dt\\
	dX^{(d)} &= \left[\mu_{1}\frac{R^{(d)}X^{(d)}}{R^{(d)}+k_{1}} - DX^{(d)} -
	\frac{\mu_{2}}{\gamma_{2}} \frac{X^{(d)}Y^{(d)}}{X^{(d)}+k_{2}}\right]\, dt
	+\varepsilon \sigma_{11}\, dB^{(d)}_{1}(t) + \varepsilon \sigma_{12}\, dB^{(d)}_{2}(t)\\
	dY^{(d)} &= \left[\mu_{2} \frac{X^{(d)}Y^{(d)}}{X^{(d)}+k_{2}} - (D+M)Y^{(d)} \right]\, dt
	+\varepsilon \sigma_{21}\, dB^{(d)}_{1}(t) + \varepsilon \sigma_{22}\, dB^{(d)}_{2}(t),
\end{align*}
\item[(B)] Noise in the intrinsic birth rates:
\begin{align*}
	dR^{(d)} &=
	\left[(R_{0}-R^{(d)})D - \frac{\mu_{1}}{\gamma_{1}} \frac{R^{(d)}X^{(d)}}{R^{(d)}+k_{1}}\right]\, dt\\
	dX^{(d)} &= \left[\mu_{1}\frac{R^{(d)}X^{(d)}}{R^{(d)}+k_{1}} - DX^{(d)} -
	\frac{\mu_{2}}{\gamma_{2}} \frac{X^{(d)}Y^{(d)}}{X^{(d)}+k_{2}}\right]\, dt\\
	&\qquad + \varepsilon \sigma_{11} \mu_{1}\frac{R^{(d)}X^{(d)}}{R^{(d)}+k_{1}}\, dB^{(d)}_{1}(t) 
		+ \varepsilon \sigma_{12} \mu_{1}\frac{R^{(d)}X^{(d)}}{R^{(d)}+k_{1}}\, dB^{(d)}_{2}(t)\\
	dY^{(d)} &= \left[\mu_{2} \frac{X^{(d)}Y^{(d)}}{X^{(d)}+k_{2}} - (D+M)Y^{(d)} \right]\, dt\\
	&\qquad + \varepsilon \sigma_{21} \mu_{2} \frac{X^{(d)}Y^{(d)}}{X^{(d)}+k_{2}}\, dB^{(d)}_{1}(t) 
		+ \varepsilon \sigma_{22} \mu_{2} \frac{X^{(d)}Y^{(d)}}{X^{(d)}+k_{2}}\, dB^{(d)}_{2}(t).
\end{align*}
\end{itemize}
We will assume that 
\[
	\E{B^{(1)}_{i}(t),B^{(2)}_{j}(t)} = \begin{cases}
		\rho_{i} t & \text{if $i=j$}\\
		0 & \text{otherwise}
	\end{cases}
\]
so that there are two independent sources of environmental noise in each community (\textit{e.g.} independent noise in temperature and in pH),  whilst these noises are correlated across communities.

\subsubsection{$M=0$}

In both cases, we have, in the notation of the previous section,
\[
	\bF(R,X,Y) = \mat{(R_{0}-R)D - \frac{\mu_{1}}{\gamma_{1}} \frac{RX}{R+k_{1}}\\
	\mu_{1}\frac{RX}{R+k_{1}} - DX -
	\frac{\mu_{2}}{\gamma_{2}} \frac{XY}{X+k_{2}}\\
	\mu_{2} \frac{XY}{X+k_{2}} - (D+M)Y},
\]
whilst the matrix $\bSigma(R,X,Y)$ takes the form 
\[
	 \bSigma_{A}(R,X,Y) = \mat{ 0 & 0 & 0\\  0 & \sigma_{11} & \sigma_{12}\\  0 & \sigma_{21} & \sigma_{22}},
\]
for model (A), and 
\[
	 \bSigma_{B}(R,X,Y) = \mat{0 & 0 & 0 \\ 0 & \sigma_{11} \mu_{1}\frac{RX}{R+k_{1}} &  \sigma_{12} \mu_{1}\frac{RX}{R+k_{1}}\\
	0 & \sigma_{21} \mu_{2} \frac{XY}{X+k_{2}} & \sigma_{22} \mu_{2} \frac{XY}{X+k_{2}}}
\]
for model (B).

The dynamical system 
\begin{equation}\label{DS}
	\frac{d}{dt} \mat{R\\X\\Y} = \bF(R,X,Y)
\end{equation}
has an equilibrium point at 
\begin{align*}
	R^{\star} &= \frac{1}{2}\left(k_{1}-R_{0}+\frac{\mu_{1}}{\gamma_{1}} \frac{X^{\star}}{D}\right)
		+ \frac{1}{2}\sqrt{\left(k_{1}-R_{0}+\frac{\mu_{1}}{\gamma_{1}} \frac{X^{\star}}{D}\right)^{2}
		+4R_{0}k_{1}}\\
	X^{\star} &= \frac{k_{2}D}{\mu_{2}-D}\\
	Y^{\star} &= \gamma_{1}\gamma_{2}(R_{0}-R^{\star})-\gamma_{2}X^{\star}
\end{align*}
for which
\[
	\ba = (\bD\bF)(R^{\star},X^{\star},Y^{\star}) 
\]
whilst $\bsigma_{A}$ and $\bsigma_{B}$ are obtained by evaluating $\bSigma_{A}$ and 
$\bSigma_{B}$ respectively at $(R^{\star},X^{\star},Y^{\star})$.

We are interested in understanding how the asymptotic correlation varies with $D$.  Some care is required, as $D$ is a bifurcation parameter for \eqref{DS}, and the linearisation is only valid for more than a very short time if $(R^{\star},X^{\star},Y^{\star})$ is an asymptotically stable fixed point.  In particular, as 
\begin{multline*}
	D \to D_{w} \defn \frac{1}{2} 
	\frac{\gamma_{1}R_{0}(\mu_{1}+\mu_{2})+\gamma_{1}\mu_{2}k_{1}+\mu_{1}k_{2}}
		{\gamma_{1}(R_{0}+k_{1})+k_2}\\
	+\sqrt{\left(\frac{\gamma_{1}R_{0}(\mu_{1}+\mu_{2})+\gamma_{1}\mu_{2}k_{1}+\mu_{1}k_{2}}
		{\gamma_{1}(R_{0}+k_{1})+k_2}\right)^{2}
		-\frac{4\gamma_{1}\mu_{1}\mu_{2}R_{0}}{\gamma_{1}(R_{0}+k_{1})+k_2}},
\end{multline*}
the system goes through a transcritical bifurcation, wherein $(R^{\star},X^{\star},Y^{\star})$ becomes unstable, corresponding to a washout of the predator.  Moreover, for a wide range of values of 
$R_{0}$, there exists  $D_{H} \in (0,D_{w})$, at which point \eqref{DS} undergoes a Hopf bifurcation: for $D_{H} < D < D_{w}$,  $(R^{\star},X^{\star},Y^{\star})$ is asymmptotically stable, whereas for
$D < D_{H}$, $(R^{\star},X^{\star},Y^{\star})$ is an unstable focus-node surrounded by a stable limit cycle (it is possible to obtain an analytic expression for $D_{H}$ using a symbolic computation package, but the expression obtained is extremely unwieldy).  Additionally, as $D$ increases through $D_{f \to n} \in (D_{H},D_{w})$, the imaginary component of  the eigenvalues of $\ba$ vanish and $(R^{\star},X^{\star},Y^{\star})$ transitions from a stable focus-node to a stable node.

\begin{figure}[h] \begin{center}
\scalebox{2.0}{\includegraphics[width=.3\textwidth]{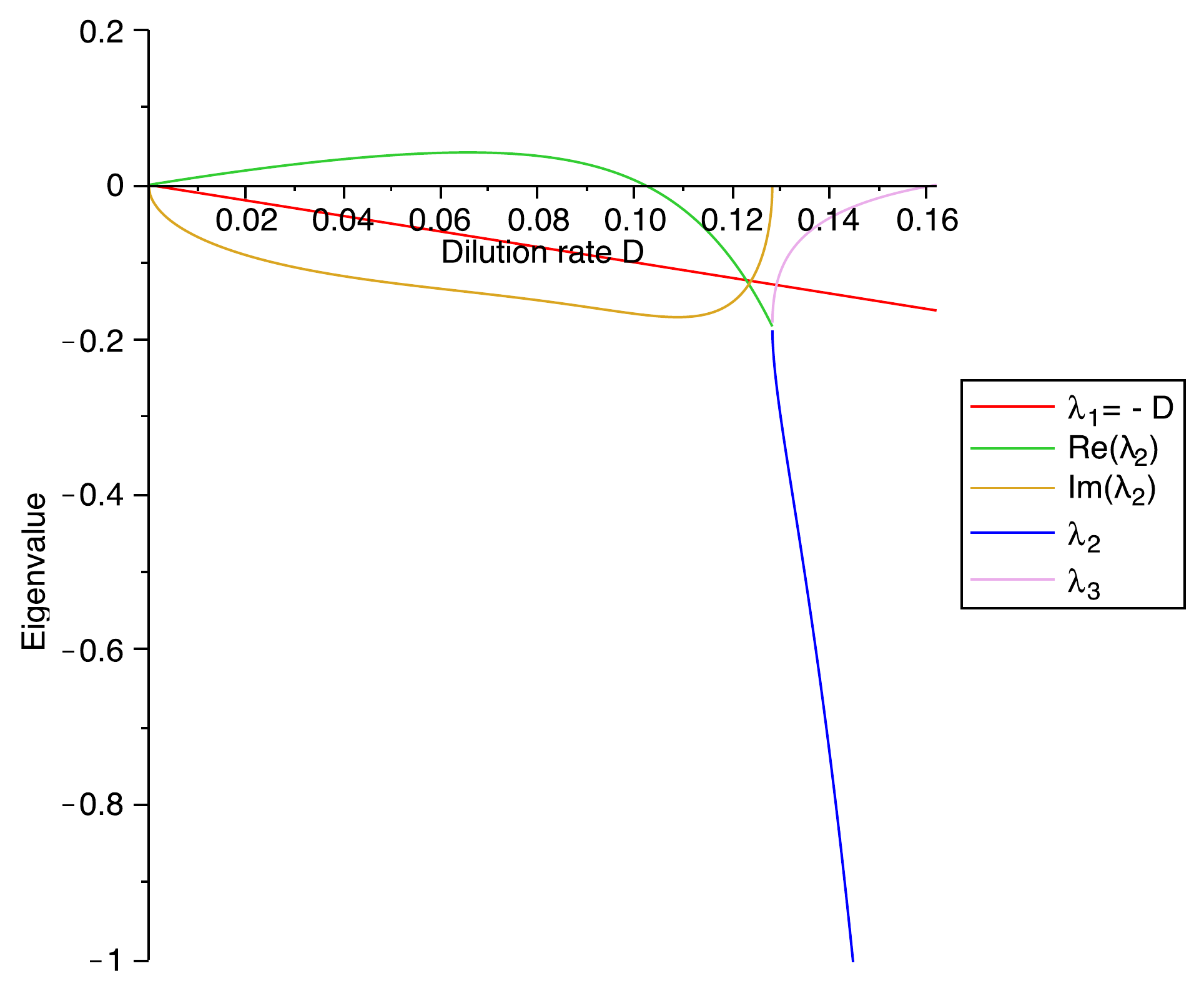}} \end{center}
\caption[]{\textbf{Eigenvalues of $\ba$.} For all values of $D$, $\ba$ has eigenvalue $\lambda_{1} = -D$.  For $0< D < D_{f \to n} \approx 0.1282994432$, $\ba$ has a pair of complex conjugate eigenvalues $\lambda_{2}=\bar{\lambda}_{3}$; the real part crosses the imaginary axis at $D_{H} \approx 0.1020177469$.  For $D_{f \to n} < D < D_{w} \approx 0.1620774695$, $\lambda_{2}$ and $\lambda_{3}$ are distinct, real, and negative.  Finally, at $D=D_{w}$, $\lambda_{3}$ vanishes as the predator washes out.}
\label{EIGENVALS}
\end{figure}

In light of the previous section, provided $\varepsilon$ is sufficiently small and $D_{H} < D < D_{w}$, we may use the results of \ref{MORAN} to approximate the correlation in the predator and prey populations across two unconnected replicate chemostats.  Again, whilst it is in principle possible to compute the correlation analytically, in practice, the expressions are too complex to be understood, and we will limit ourselves to numerically illustrating the possible qualitative behaviours for a representative set of parameters, given in Table \ref{PARAMS}.  For these parameters, we have $D_{H} \approx 0.1020177469$, $D_{f \to n} \approx 0.1282994432$, $D_{w} \approx 0.1620774695$ (see Figure \ref{EIGENVALS}).

\newcolumntype{L}{>{$}l<{$}} \newcolumntype{C}{>{$}c<{$}}

\begin{longtable}{C@{\extracolsep{10mm}} l @{\extracolsep{10mm}} C}
\caption{Parameter Values}\\
\text{Parameter} & Interpretation & \text{Typical value}\\
\hline
R_{0}		& inflow concentration				& 100\\
\mu_{1}		& maximal growth rate (prey)			& 0.5\\
\mu_{2}		& maximal growth rate (predator)		& 0.2\\
k_{1}			& half saturation constant (prey)		& 8\\
k_{2}			&  half saturation constant (predator)	& 9\\
\gamma_{1}	& yield (prey)						& 0.4\\
\gamma_{2}	& yield (predator)					& 0.6\label{PARAMS}
\end{longtable}

\begin{figure} \begin{center}
\mbox{
\subfigure[]{\includegraphics[width=.45\textwidth]{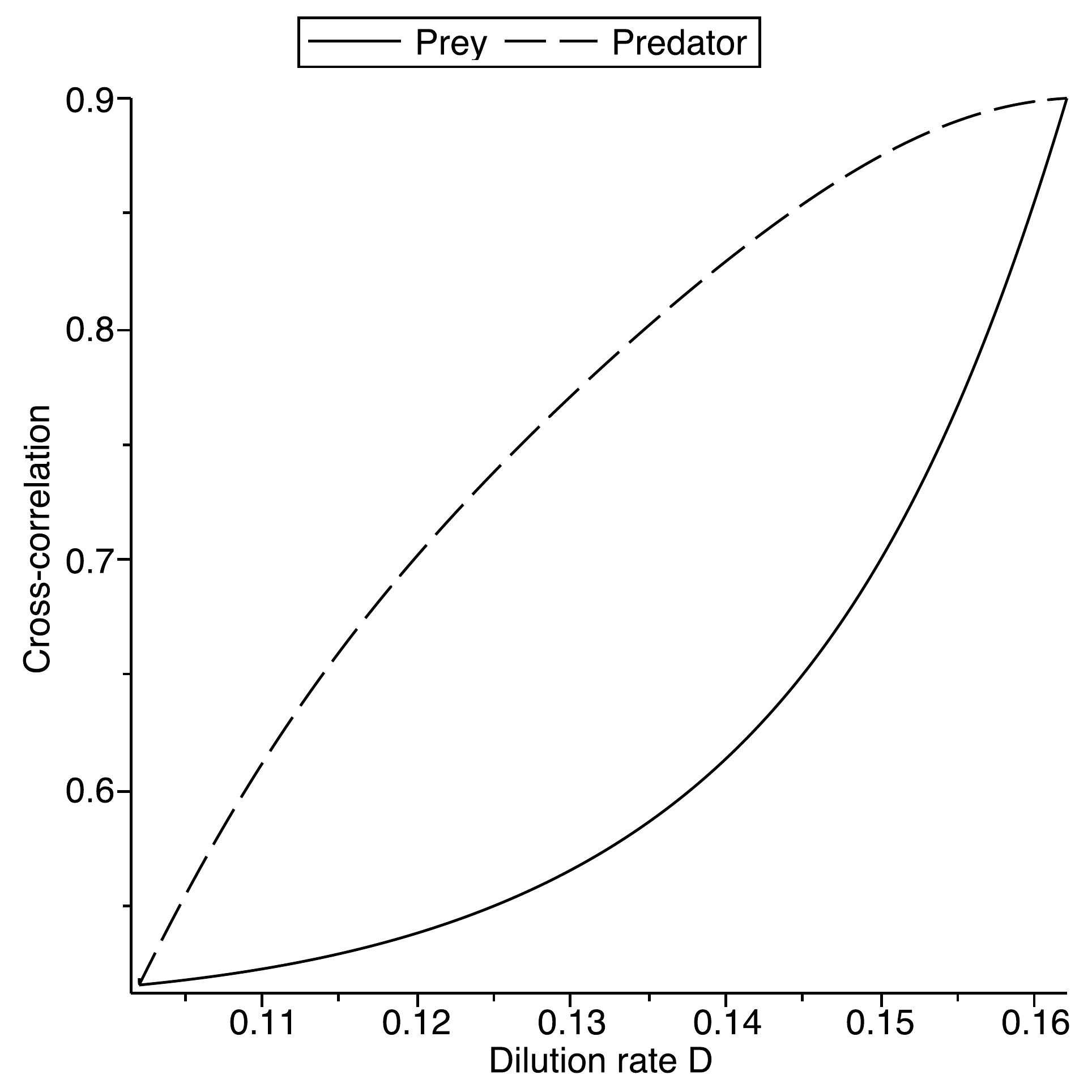}}
\subfigure[]{\includegraphics[width=.45\textwidth]{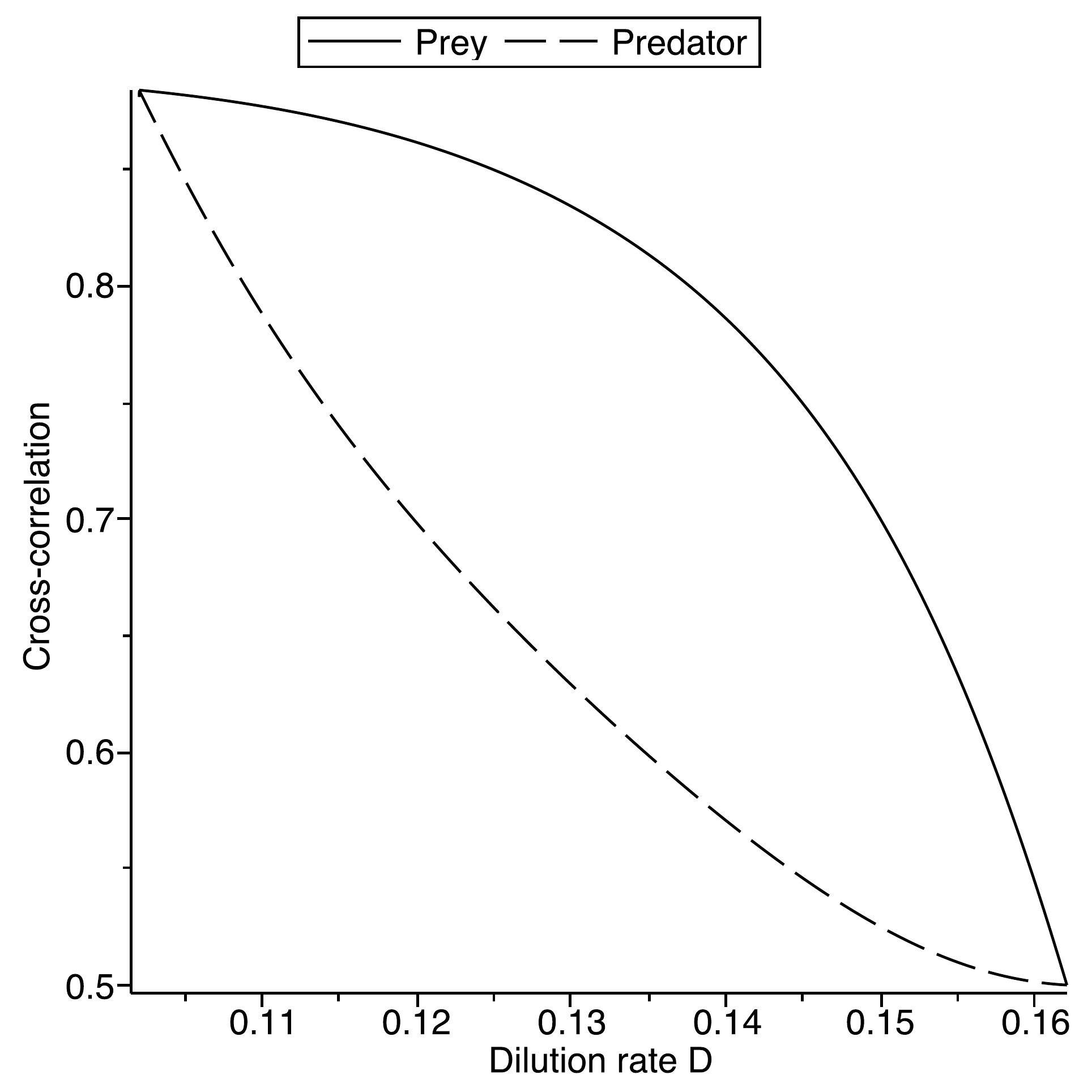}}}
\end{center} \caption[]{Correlation in predator and prey populations for model (A) with $M=0$, $\sigma_{11} = 0.5$, $\sigma_{12} = 0.3$, $\sigma_{21} = 0.05$, $\sigma_{22} = 0.1$, (a) $\rho_{1} = 0.5$, $\rho_{2} = 0.9$, and (b) $\rho_{1} = 0.9$, $\rho_{2} = 0.5$.} \label{CORRA}
\end{figure}

\begin{figure} \begin{center}
\mbox{
\subfigure[]{\includegraphics[width=.45\textwidth]{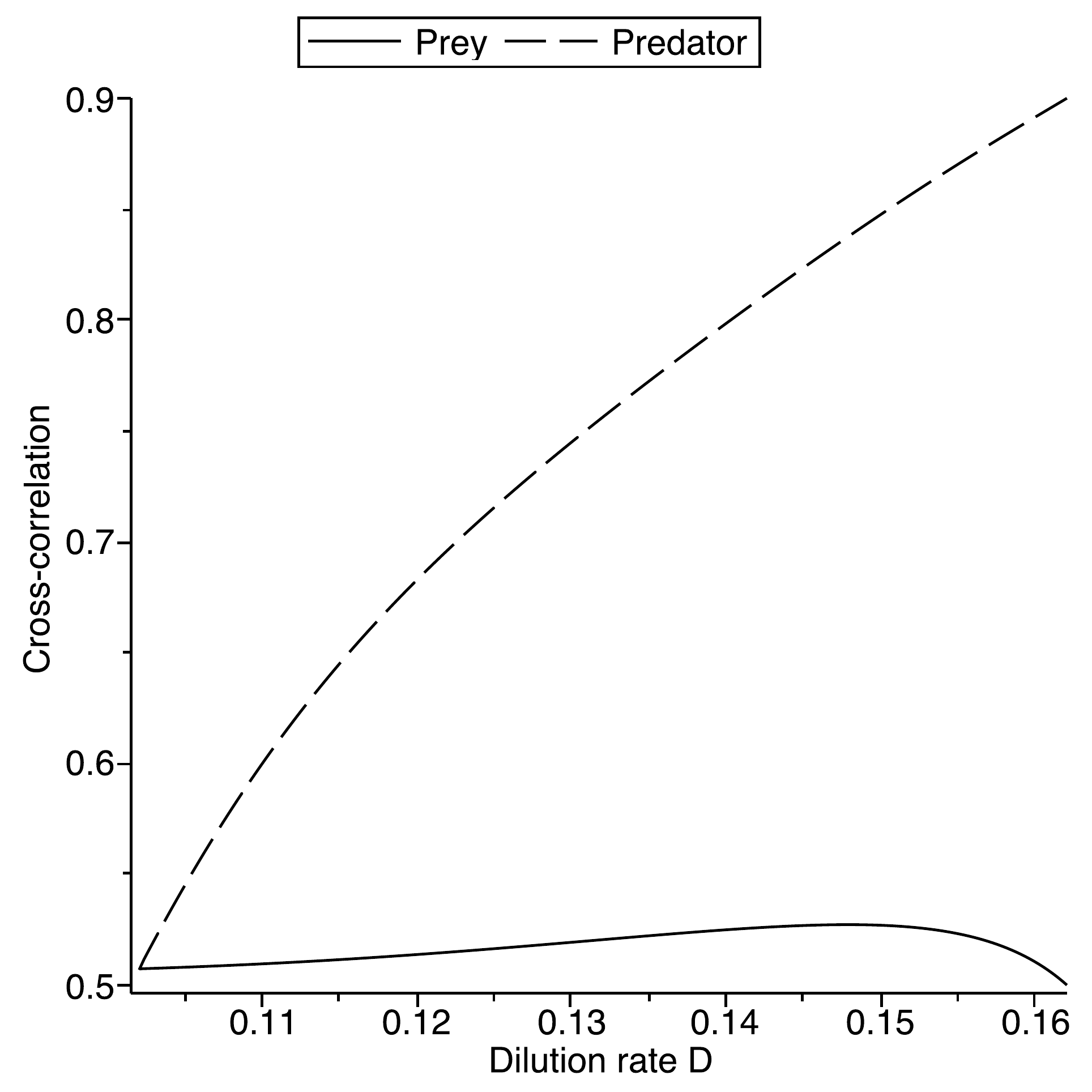}}
\subfigure[]{\includegraphics[width=.45\textwidth]{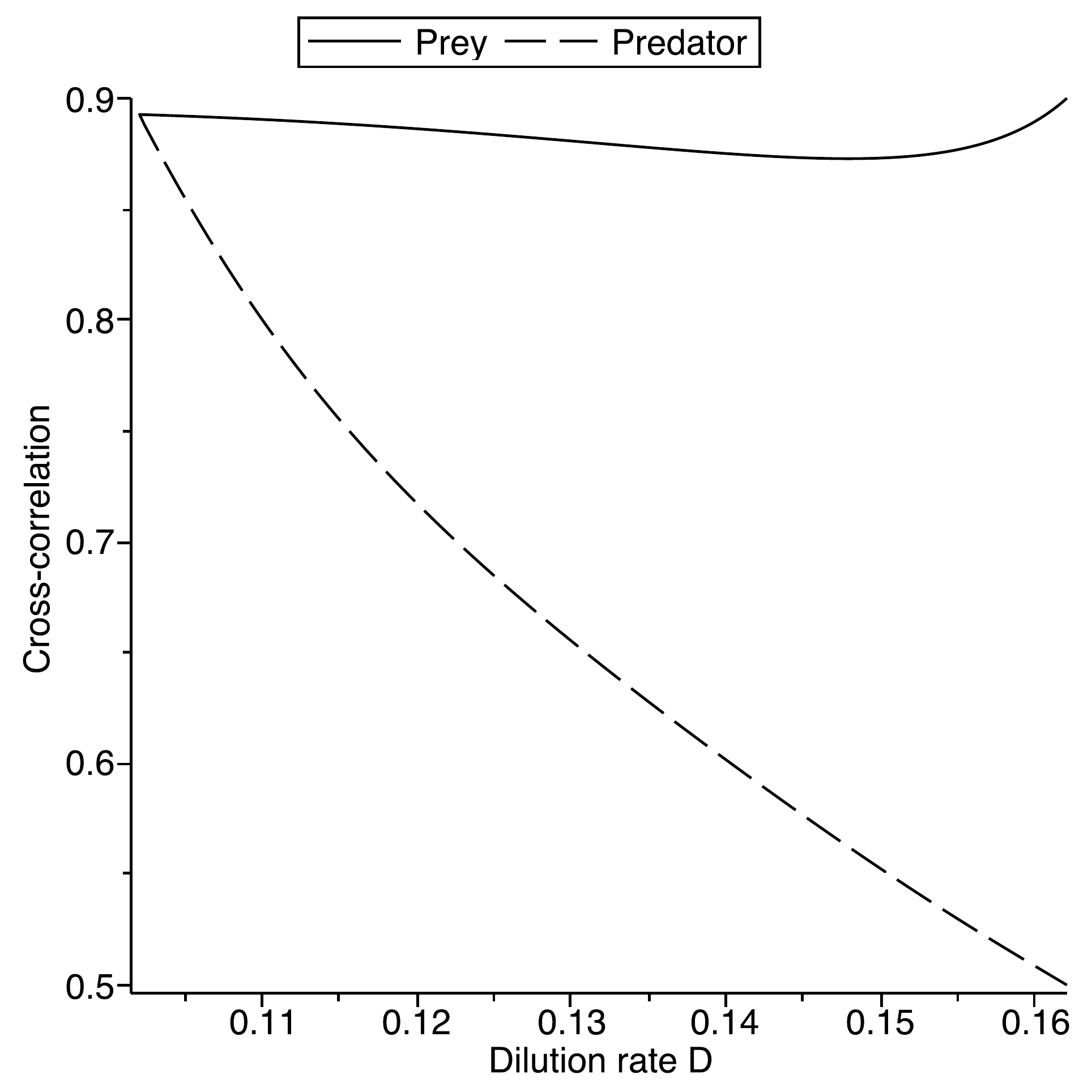}}}
\end{center} \caption[]{Correlation in predator and prey populations for model (B) with $M=0$, $\sigma_{11} = 0.5$, $\sigma_{12} = 0.3$, $\sigma_{21} = 0.05$, $\sigma_{22} = 0.1$, (a) $\rho_{1} = 0.5$, $\rho_{2} = 0.9$, and (b) $\rho_{1} = 0.9$, $\rho_{2} = 0.5$.} \label{CORRB}
\end{figure}

From Figures \ref{CORRA} and \ref{CORRB}, we see that provided $\rho_{1} \neq \rho_{2}$ (\textit{cf.} Equation \eqref{EQRHO}), the correlation in both predator and prey is in general non-monotonic in $D$.  Moreover, as the various examples illustrate, contrary to the result in one dimension, the degree of correlation, and indeed its qualitative properties, is quite sensitive to the form taken by the noise.  

Of particular interest, in both models (A) and (B), fluctuations in predator and prey become equally correlated at the Hopf bifurcation ($D=D_{H}$).  For model (A), the same occurs at the transcritical bifurcation, $D=D_{w}$ (Figure \ref{CORRA}).  Indeed, both approach the same correlation, 0.9,  as the exogenous noise acting on the prey, counterintuitively at precisely the value of $D$ at which the predator disappears.   We explain these phenomena in the next section. 

\subsubsection{$M > 0$}

In this section, we briefly illustrate the sensitivity to the form of the model, 
by considering consequences of introducing intrinsic mortality in the predator 
\textit{i.e.} taking $M > 0$.  This leaves the qualitative dynamics unchanged, 
but enlarges the region in which the interior equilibrium is stable \citep{Nisbet83}.  
Indeed, taking $M=0.1$, and otherwise using the parameter values given in Table \ref{PARAMS}, 
we see that there is no Hopf bifurcation, whilst $D_{w} \approx 0.06291670082$, so that the
 linearisation is applicable for $0<D<D_{w}$.

\begin{figure} \begin{center}
\mbox{
\subfigure[]{\includegraphics[width=.45\textwidth]{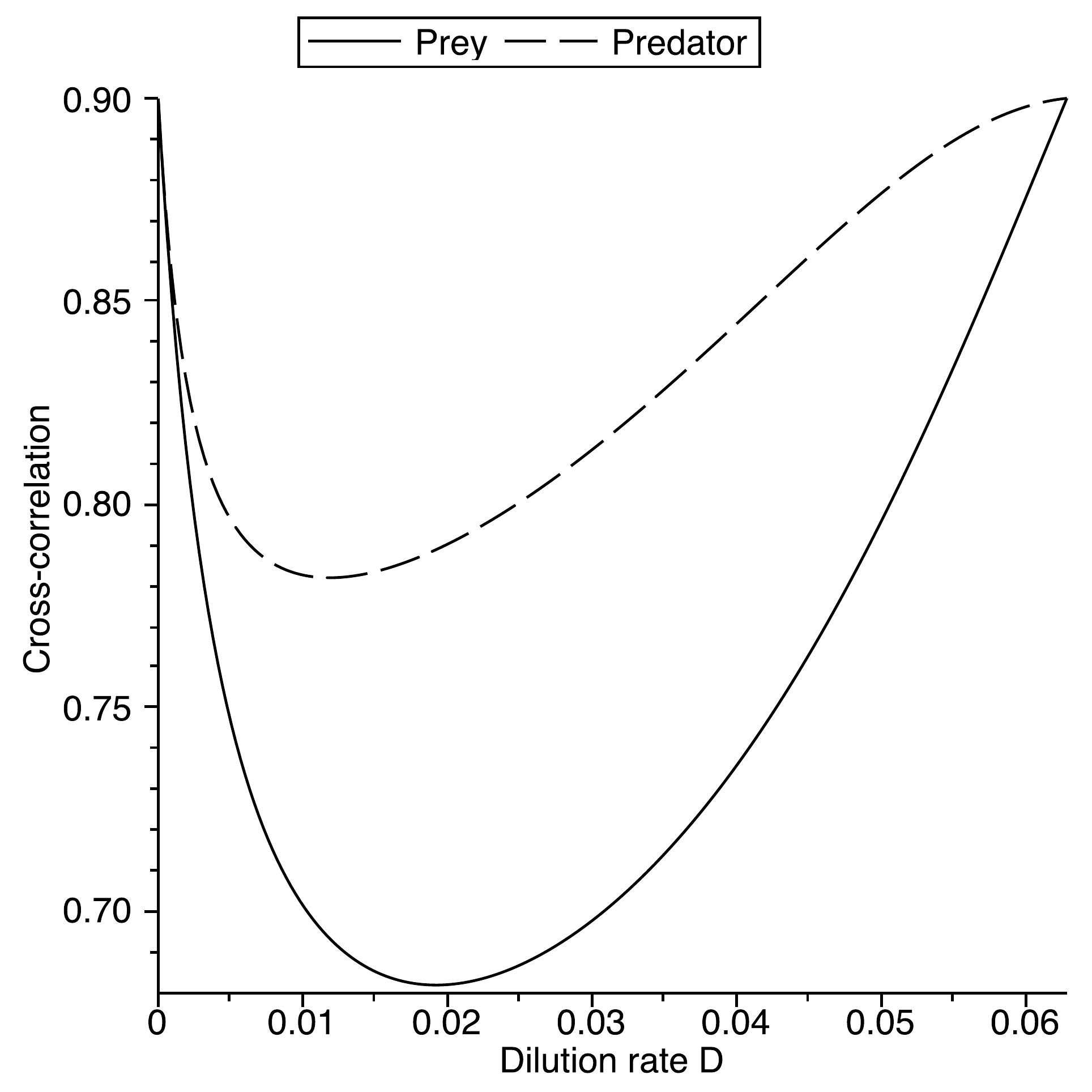}}
\subfigure[]{\includegraphics[width=.45\textwidth]{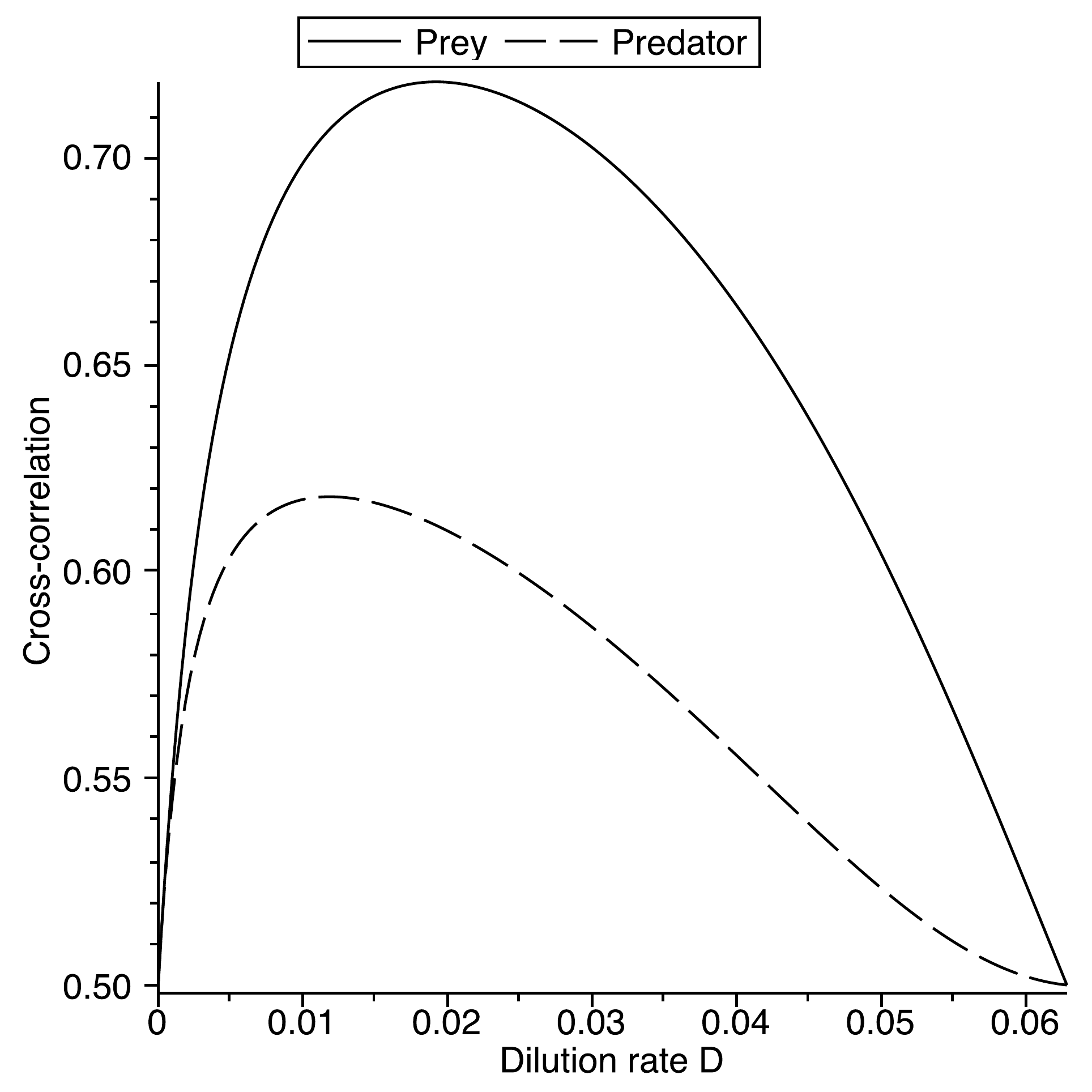}}}
\end{center} \caption[]{Correlation in predator and prey populations for model (A) with $M=0.1$, $\sigma_{11} = 0.5$, $\sigma_{12} = 0.3$, $\sigma_{21} = 0.05$, $\sigma_{22} = 0.1$, (a) $\rho_{1} = 0.5$, $\rho_{2} = 0.9$, and (b) $\rho_{1} = 0.9$, $\rho_{2} = 0.5$.} \label{CORRAM}
\end{figure}

\begin{figure} \begin{center}
\mbox{
\subfigure[]{\includegraphics[width=.45\textwidth]{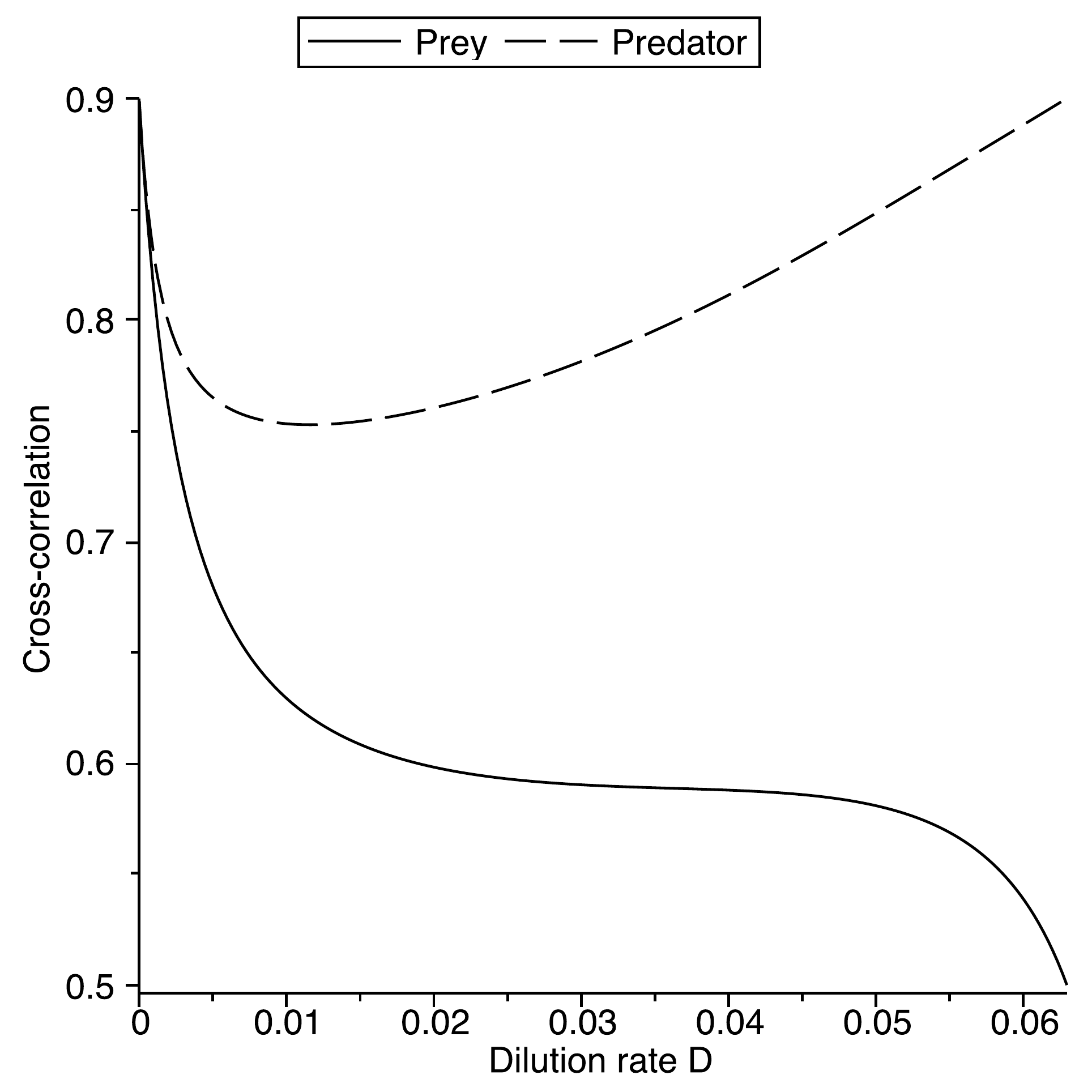}}
\subfigure[]{\includegraphics[width=.45\textwidth]{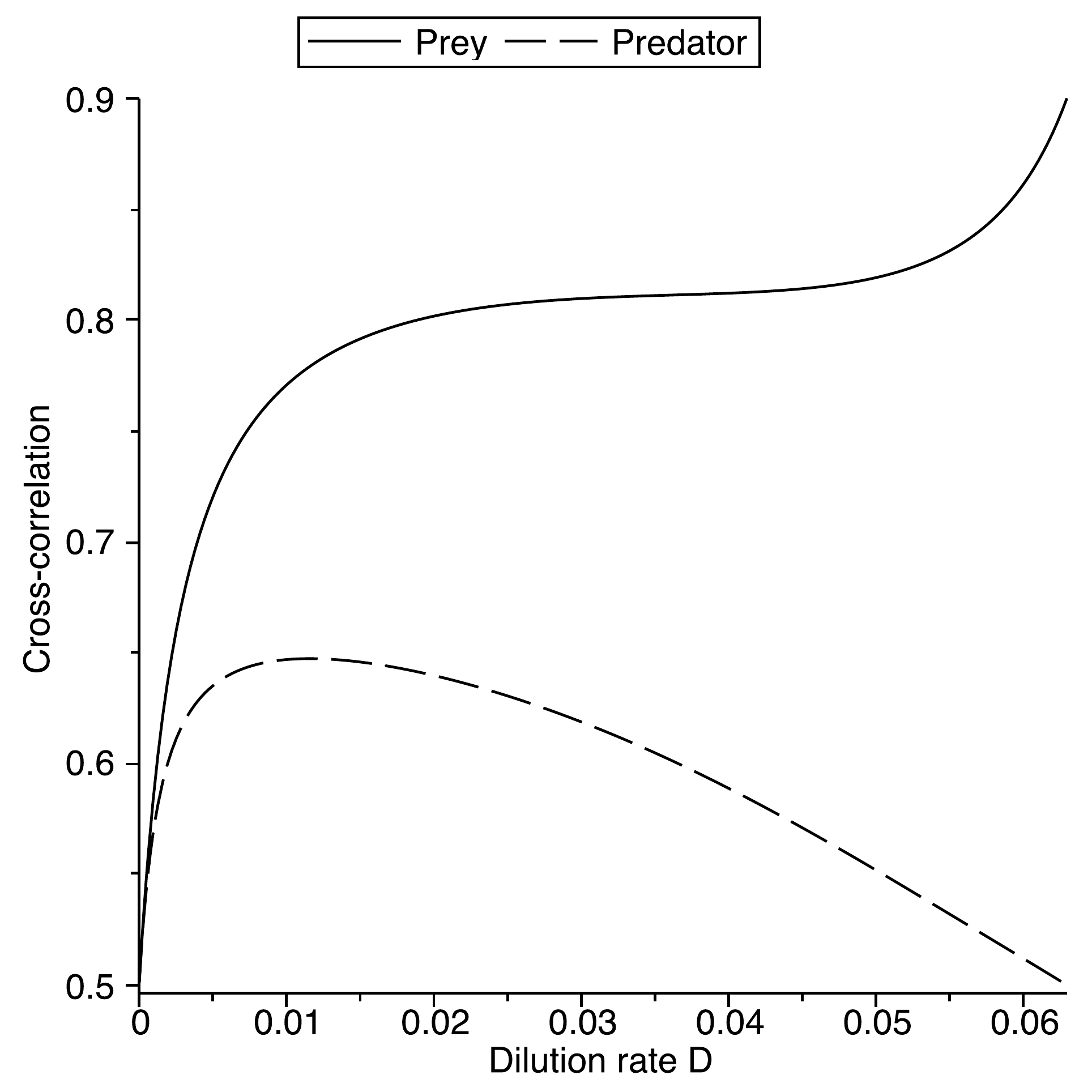}}}
\end{center} \caption[]{Correlation in predator and prey populations for model (B) with $M=0.1$, $\sigma_{11} = 0.5$, $\sigma_{12} = 0.3$, $\sigma_{21} = 0.05$, $\sigma_{22} = 0.1$, (a) $\rho_{1} = 0.5$, $\rho_{2} = 0.9$, and (b) $\rho_{1} = 0.9$, $\rho_{2} = 0.5$.} \label{CORRBM}
\end{figure}

\subsection{Bifurcations and Correlation}

To explain the correlation phenomena observed as $D$ approaches bifurcation points, we return to the general framework developed in \ref{ShortLong}.  For simplicity in our exposition, we will assume that the matrix $\ba$ capturing the linearised dynamics has $m$ eigenvalues of multiplicity one, $\lambda_{1},\ldots,\lambda_{m}$, with corresponding eigenvectors $\bdf^{(1)},\ldots,\bdf^{(m)}$.   Without loss of generality, we assume that the eigenvalues of $\ba$ are ordered so that  if $\lambda_{j}$ is a non-real eigenvalue (\textit{i.e.} $\lambda_{j} \in \C-\R$), then $\lambda_{j+1}=\overline{\lambda_{j}}$.

In general, bifurcations will occur at points where the real part of some eigenvalue vanishes, \textit{i.e.} $\Re(\lambda_{i}) = 0$ for some $i$. As we have already observed in our explicit chemostat model, fluctuations in the numbers both predator and prey become equally correlated at these bifurcation points.  As we shall show below, this is a general phenomenon.

We begin by observing that $\ba \otimes I_{m \times m} + I_{m \times m} \otimes \ba$ has eigenvalues $\lambda_{i}+\lambda_{j}$ corresponding to eigenvectors 
\[
	F^{(i,j)}=\text{vec}(\bdf^{(i)} \otimes \bdf^{(j)}),
\]
($\otimes$ denotes the Kronecker product) for $i,j = 1,\ldots,m$.  In particular, $2\lambda_{i}$ is an eigenvalue of $\ba \otimes I_{m \times m} + I_{m \times m} \otimes \ba$ for all $i$, and, if $\lambda_{j} \in \C-\R$ is an eigenvalue of $\ba$, then $\lambda_{j+1} =\overline{\lambda_{j}}$ is as well, so 
$2\Re(\lambda_{j})=\lambda_{j}+\lambda_{j+1}$ is also an eigenvalue of 
$\ba \otimes I_{m \times m} + I_{m \times m} \otimes \ba$.  

The linear system \eqref{TENEQ} can then be formally solved using these eigenfunction expansions.  Let $\{\check{\bdf}^{(i)}\}_{i=1}^{m}$ be a basis dual to $\{\bdf^{(i)}\}_{i=1}^{m}$:
\[
	\check{\bdf}^{(i)} \cdot \bdf^{(j)} = \begin{cases} 
		1 & \text{if $i=j$}\\
		0 & \text{otherwise.}
	\end{cases}
\]
Then $\{\check{F}^{(i,j)}\}_{i,j=1}^{m}$ is the corresponding dual basis to $\{F^{(i,j)}\}_{i,j=1}^{m}$ and we may write 
\begin{gather*}
	C = \sum_{i,j=1}^{m} \ip{C}{\check{F}^{(i,j)}} F^{(i,j)},\\
	\sum_{k=1}^{n} \rho_{k} \Sigma^{(k)} 
	= \sum_{k=1}^{n} \rho_{k} \sum_{i,j=1}^{m} \ip{\Sigma^{(k)}}{\check{F}^{(i,j)}} F^{(i,j)},\\
\intertext{and}
	(\ba \otimes I_{m \times m} + I_{m \times m} \otimes \ba) C 
	= \sum_{i,j=1}^{m} (\lambda_{i}+\lambda_{j}) \ip{C}{\check{F}^{(i,j)}} F^{(i,j)},
\end{gather*}
where $\Sigma^{(k)}$ is defined as in \eqref{DEFCSIGMA}.  Then, matching the coefficients of like basis vectors, we have
\[
	\ip{C}{\check{F}^{(i,j)}} 
	= \frac{\sum_{k=1}^{n} \rho_{k} \ip{\Sigma^{(k)}}{\check{F}^{(i,j)}}}{\lambda_{i}+\lambda_{j}}.
\]

Now, recall that $C = \text{vec}(\bc)$, so that, if we let $E^{(i,j)} = \text{vec}(\be^{(i,j)})$
(recall, $\be^{(i,j)}$  is the $m \times m$ matrix with $ij$\textsuperscript{th} entry equal to one, and all other entries equal to zero, so  $E^{(i,j)}$ is the $(i-1)m+j$\textsuperscript{th} standard basis vector in $\R^{m^{2}}$), then
\[
	c_{ij} = \ip{C}{E^{(i,j)}}
	= \sum_{p,q=1}^{m} \frac{\sum_{k=1}^{n} \rho_{k} \ip{\Sigma^{(k)}}{\check{F}^{(p,q)}}}
	{\lambda_{p}+\lambda_{q}} \ip{F^{(p,q)}}{E^{(i,j)}},
\]
and, in a similar fashion,
\[
	v_{ij} = \ip{V}{E^{(i,j)}}
	=  \sum_{p,q=1}^{m} \frac{\sum_{k=1}^{n} \ip{\Sigma^{(k)}}{\check{F}^{(p,q)}}}{\lambda_{p}
	+\lambda_{q}} \ip{F^{(p,q)}}{E^{(i,j)}},
\]
We thus have 
\begin{align}\label{EIGENCOR}
	\lim_{t \to \infty} \text{corr}(X_{1i}(t),X_{2i}(t)) &= \frac{c_{ii}}{v_{ii}}\\
	&= \frac{\sum_{p,q=1}^{m} 
	\frac{\sum_{k=1}^{n} \rho_{k} \ip{\Sigma^{(k)}}{\check{F}^{(p,q)}}}{\lambda_{p}
	+\lambda_{q}} \ip{F^{(p,q)}}{E^{(i,i)}}}{\sum_{p,q=1}^{m} \frac{\sum_{k=1}^{n} 	
	\ip{\Sigma^{(k)}}{\check{F}^{(p,q)}}}{\lambda_{p}+\lambda_{q}} \ip{F^{(p,q)}}{E^{(i,i)}}}.
\end{align}

Now, consider the situation where the eigenvalues vary with some bifurcation parameter, $\beta$ and suppose that, without loss of generality,
\[
	\Re(\lambda_{1}) \to 0
\]
as $\beta \to \beta^{\star}$ for some fixed constant, $\beta^{\star}$.   We consider two cases,
\begin{itemize}
\item[(i)] $\lambda_{1} \in \R$: Then, provided $\ip{\Sigma^{(k)}}{\check{F}^{(1,1)}} \neq 0$, multiplying the numerator and denominator of \eqref{EIGENCOR} by $\lambda_{1}$ and simplifying, we have
\begin{equation}\label{CORREVR}
	\frac{c_{ii}}{v_{ii}} = \frac{\sum_{k=1}^{n} \rho_{k} \ip{\Sigma^{(k)}}{\check{F}^{(1,1)}}}
	{\sum_{k=1}^{n} \ip{\Sigma^{(k)}}{\check{F}^{(1,1)}}} + \BigO{\lambda_{1}},
\end{equation}
\item[(ii)] $\lambda_{1}, \lambda_{2} = \overline{\lambda_{1}} \in \C-\R$:
Then, we have $\lambda_{1} + \lambda_{2} =\Re(\lambda_{1})$, so that, 
assuming at least one of  $\ip{\Sigma^{(k)}}{\check{F}^{(1,2)}}$ or $\ip{\Sigma^{(k)}}{\check{F}^{(2,1)}}$ is non-zero, 
\begin{equation}\label{CORREVC}
	\frac{c_{ii}}{v_{ii}} 
	=\frac{\sum_{k=1}^{n} \rho_{k} \left(\ip{\Sigma^{(k)}}{\check{F}^{(1,2)}}\ip{F^{(1,2)}}{E^{(i,i)}}
	+\ip{\Sigma^{(k)}}{\check{F}^{(2,1)}}\ip{F^{(2,1)}}{E^{(i,i)}}\right)}
	{\sum_{k=1}^{n} \ip{\Sigma^{(k)}}{\check{F}^{(1,1)}}\ip{F^{(1,2)}}{E^{(i,i)}}
	+\ip{\Sigma^{(k)}}{\check{F}^{(2,1)}}\ip{F^{(2,1)}}{E^{(i,i)}}} + \BigO{\Re(\lambda_{1})}.
\end{equation}
But 
\[	
	F^{(1,2)} \cdot E^{(i,i)} = f^{(1)}_{i} f^{(2)}_{i} = \abs{f^{(1)}_{i}}^{2},
\]
and similarly $\ip{F^{(2,1)}}{E^{(i,i)}} = \abs{f^{(1)}_{i}}^{2}$, so 
\begin{align*}
	\frac{c_{ii}}{v_{ii}} &=\frac{\sum_{k=1}^{n} \rho_{k} \left(\ip{\Sigma^{(k)}}{\check{F}^{(1,2)}}
	+\ip{\Sigma^{(k)}}{\check{F}^{(2,1)}}\right)}
	{\sum_{k=1}^{n} \ip{\Sigma^{(k)}}{\check{F}^{(1,2)}}
	+\ip{\Sigma^{(k)}}{\check{F}^{(2,1)}}} + \BigO{\Re(\lambda_{1})}\\
	&= \frac{\sum_{k=1}^{n} \rho_{k} \Re\ip{\Sigma^{(k)}}{\check{F}^{(1,2)}}}
	{\sum_{k=1}^{n} \Re\ip{\Sigma^{(k)}}{\check{F}^{(1,2)}}} + \BigO{\Re(\lambda_{1})}\\
\end{align*}
\end{itemize}

Thus, as $\Re(\lambda_{1}) \to 0$, the correlation $\frac{c_{ii}}{v_{ii}}$ becomes independent of the species type $i$, as we observed above.  We also observe that the correlation, although determined by the vanishing of the real part of the eigenvalue, does not otherwise depend on the eigenvalue, whilst the contribution of the various $\rho_{k}$ to the total correlation is proportional to the magnitude of the projection of the vectors $\Sigma^{(k)}$ onto the eigenvector $F^{(1,1)}$ or the eigenvectors $F^{(1,2)}$, $F^{(2,1)}$ in the real and complex cases respectively.  

When $\Sigma^{(k)}$ is proportional to $E^{(k,k)}$ for all $k$, say $\Sigma^{(k)} = \sigma_{k} E^{(k,k)}$, we can further simplify the expressions above.  When $\lambda_{1} \in \R$, we have
\[	
	\check{F}^{(1,1)} \cdot E^{(k,k)} = \abs{\check{\bdf}^{(1)}_{k}}^{2},
\]
whilst $\check{F}^{(1,2)} \cdot E^{(k,k)} = \abs{\check{f}^{(1)}_{k}}^{2}$ when $\lambda_{1} \in \C-\R$, as we observed above.  Then, in both cases $(i)$ and $(ii)$, we have
\begin{equation}\label{CORRSIMP}
	\frac{c_{ii}}{v_{ii}} 
	= \frac{\sum_{k=1}^{n} \rho_{k} \sigma_{k}^{2} \abs{\check{f}^{(1)}_{k}}^{2}}
	{\sum_{k=1}^{n} \sigma_{k}^{2} \abs{\check{f}^{(1)}_{k}}^{2}} + \BigO{\lambda_{1}},
\end{equation}

\subsubsection{Application to the Predator-Prey Chemostat Model}
\label{appapp}
\begin{figure} \begin{center}
\mbox{
\subfigure[$\bdf^{(1)}$]{\includegraphics[width=.3\textwidth]{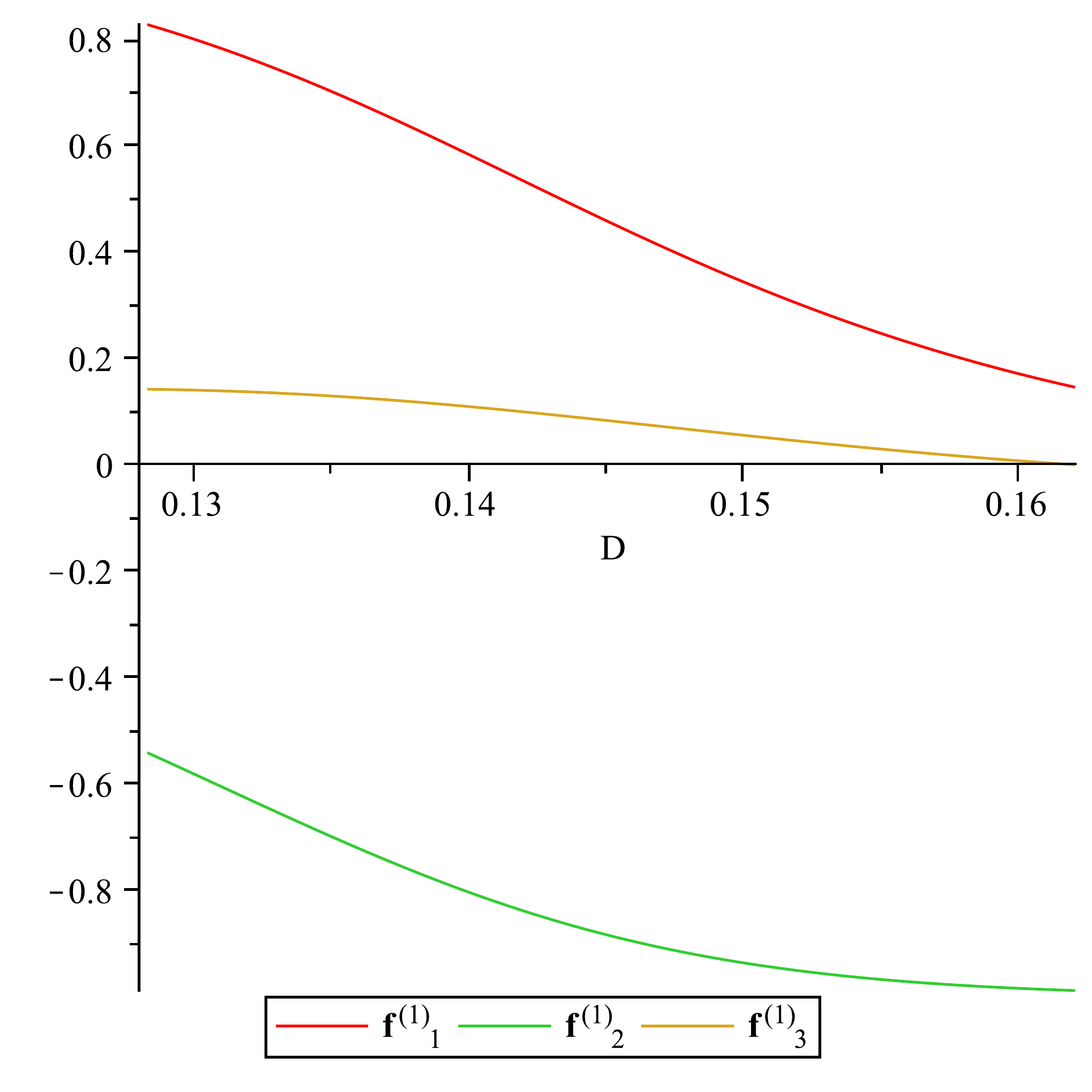}}
\subfigure[$\bdf^{(2)}$]{\includegraphics[width=.3\textwidth]{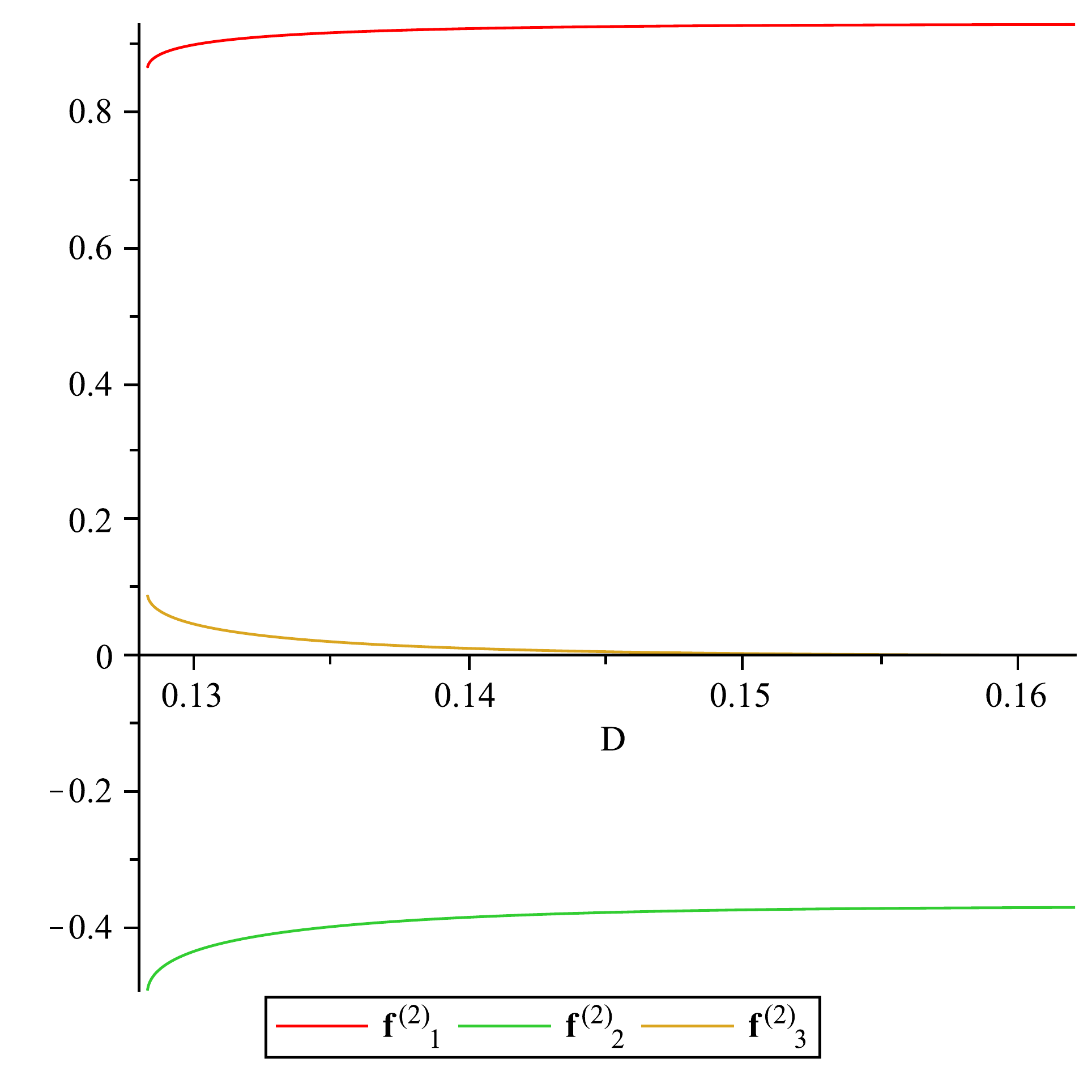}}
\subfigure[$\bdf^{(3)}$]{\includegraphics[width=.3\textwidth]{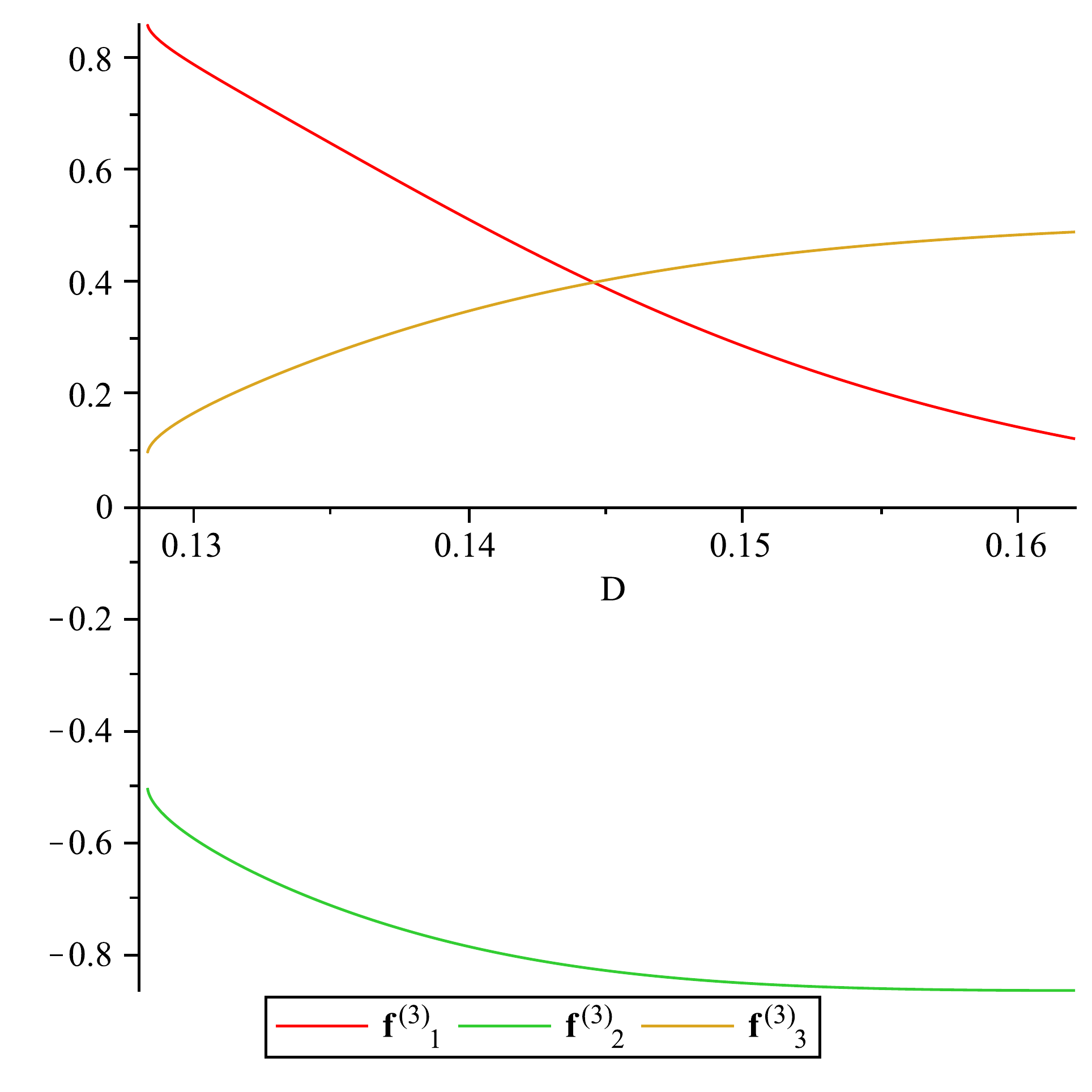}}}
\end{center} \caption[]{Components of the eigenvectors of $\ba$ for model (A) as a function of $D$,} \label{EIGENVECTOR}
\end{figure}

Consider model $(A)$, with linear noise.  Then 
\begin{gather*}
	\lambda_{1} \to -D_{w}\\
	\lambda_{2} \to -\frac{(D_{w}-\mu_{1})(R_{0}(\mu_{1}-D_{w})-k_{1}D_{w})}{k_{1}\mu_{1}}\\
\intertext{and}
	\lambda_{3} \to 0.
\end{gather*}
as $D \to D_{w}$.  Let $\bdf^{(1)},\bdf^{(2)},\bdf^{(3)}$ be the corresponding eigenvectors. Then,
at $D=D_{w}$, $f^{(1)}_{3}=f^{(2)}_{3}=0$ (Figure \ref{EIGENVECTOR}) and thus $\check{f}^{(3)}_{1}=\check{f}^{(3)}_{2}=0$ whilst all other components of both eigenvectors and eigenvalues are non-zero.  In particular, using the results above, we have
\[
	\frac{c_{ii}}{v_{ii}} 
	= \rho_{3} + \BigO{\lambda_{3}} 
\]
\textit{i.e.} for all $i$, $\frac{c_{ii}}{v_{ii}} \to \rho_{3}$ as $D \to D_{w}$ as observed in  observed in Figure \ref{CORRA}.

In a similar manner, we can obtain the correlations for models (A) and (B) as $D \to D_{H}$, $D \to D_{w}$, or $D \to 0$.  Some care is required in analyzing model (B) as $D \to D_{w}$, as 
\[
	\bsigma_{B} = \bSigma_{B}(R^{\star},X^{\star},Y^{\star}) \to 0 
\]
and thus $\ip{\Sigma^{(k)}}{\check{F}^{(1,1)}}=0$ for all $k$.  In particular, \eqref{CORREVR} no longer holds, all terms in \eqref{EIGENCOR} must be considered, and the correlation is no longer type-independent.

\subsection{Limitations of Linearisation}

This argument, however, also needs to be considered carefully; Blagove{\v s}{\v c}enskii and Freidlin's proof of the error estimate \eqref{LINEARERROR} is obtained via Gronwall's inequality, which gives an upper bound on the approximating error that is growing exponentially in time, with rate proportional to $\BigO{\varepsilon^2}$; thus we have no \textit{a priori} reason to believe that the populations will remain correlated for times longer than $\BigO{\abs{\ln \varepsilon}}$.

Indeed, consider the following example with multiplicative noise
\begin{equation}\label{GBM}
	dX_i(t) = a X_i(t)\, dt + b X_i(t)\, dB_i(t).
\end{equation}
When $b$ is small, this process has \eqref{OU} as its linear approximation.  However,  unlike 
\eqref{OU}, here the noise has a mechanistic intepretation: this process emerges as a limit of the 
linear birth-death process in which the birth and death rates are assumed to experience 
uncorrelated fluctuations about fixed mean values.  As before, we have 
\[
	\E{X_i(t)} = e^{at}X_i(0).
\]
We may determine the variance and the covariance of the processes using It\^o's product rule:
\begin{equation}\label{I1}
	dX^2_i(t) = 2X_i(t)\, dX_i(t) + d[X_i](t),
\end{equation}
whilst 
\begin{equation}\label{I2}
	dX_1(t)X_2(t) = X_1(t)\, dX_2(t) + X_2(t)\, dX_1(t) + d[X_1,X_2](t),
\end{equation}

where the quadratic variations and covariation are given by 
\[
	[X_i](t) = \int_{0}^t b^2 X^2_i(s)\, d[B_i](s) =  \int_{0}^{t} b^2 X^2_i(s)\, ds
\]
and
\[
	[X_1,X_2](t) = \int_{0}^{t} b^2 X_1(s) X_2(s)\, d[B_1,B_2](s) 
	=\int_{0}^{t} b^2 X_1(s) X_2(s) \rho\, ds
\]
respectively.   Taking expectations in \eqref{I1} and \eqref{I2} yield ODEs for the second moments,
\[
	\frac{d}{dt} \E{X^2_i(t)} = (2a+b^2) \E{X^2_i(t)}
\]
and
\[
	\frac{d}{dt} \E{X_1(t)X_2(t)} = (2a+\rho b^2) \E{X_1(t)X_2(t)},
\]
which may be readily solved and used in conjunction with the expression for the mean to obtain an 
expression for the correlation of the two populations:
\[
	\text{corr}(X_1(t),X_2(t)) = \frac{e^{\rho b^2 t}-1}{e^{b^2 t}-1}.
\]
Thus, 
\[
	\lim_{t \to 0} \text{corr}(X_1(t),X_2(t)) = \rho,
\]
but for $t \gg 0$,
\begin{equation}\label{ED}
	\text{corr}(X_1(t),X_2(t)) \sim e^{(\rho-1)b^2 t},
\end{equation}
which decays exponentially in time with a half-life of $\BigO{b^{-2}}$.  Thus, the small-noise 
linear approximation and Moran's theorem successfully predicts the correlation of the populations 
over short time periods, they fail to show the eventual decay in correlation that arises from a more 
mechanistic model of noise.

We also remark, without proof, that the exponential decay in correlation strength, is even more 
rapid,
\[
	\text{corr}(X_1(t),X_2(t)) \sim e^{\frac{2}{3} (\rho-1) b^2 t^3}
\]
when we consider a model with correlated noise in the birth rates
\begin{align*}
	dX_i(t) &= a(t) X_i(t)\, dt\\
	da(t) &= b\, dB_i(t).
\end{align*}
  
Note, however, that in both of these models, the population size grows exponentially on average.  By contrast, in the models considered in \ref{PPCM}, the populations rapidly approach the neighbourhood of an asymptotically stable fixed point $\bx^{\star}$ and subsequently fluctuate about that fixed point.   In this case, we expect the linearization to be valid for considerably longer than would be anticipated using Gronwall's inequality, which is a very coarse estimate.  Indeed, the arguments in \citet{Barbour76} may be adapted to show that, provided $C_{\varepsilon} = o\left(\varepsilon^{-\frac{1}{4}}\right)$,  the expected time to leave a neighbourhood of $\bx^{\star}$ of radius $\varepsilon C_{\varepsilon}$ is exponentially distributed with rate $\BigO{C_{\varepsilon}^{2}}$, so that, provided that the magnitude of the noise, $\varepsilon$, is small, the linearization gives an accurate estimate of the correlation over observed time-scales.

Comparing these examples, it becomes clear that when searching for the origins and maintenance of population synchrony in correlated environmental noise, it is vitally important to derive a form for the noise from a mechanistic description of the origin of the noise and its effects on the relevant vital rates.







\end{document}